\documentclass[utf8]{frontiersSCNS_mod} 
\usepackage{silence}
\usepackage{url,lineno,microtype}
\usepackage{amssymb}
\usepackage{mathtools}
\usepackage[onehalfspacing]{setspace}
\usepackage[acronym, toc, xindy, noredefwarn]{glossaries}
\usepackage{lmodern}
\usepackage{todonotes}

\usepackage{hyperref}

\usepackage{cleveref}
\usepackage{desclist}
\usepackage[per-mode = symbol]{siunitx}
\usepackage{tabularx}
\usepackage{tabulary}
\usepackage{booktabs}
\usepackage{multirow}
\usepackage{xfrac}

\newacronym{abs}{ABS}{anti-lock braking system}
\newacronym{auv}{AUV}{autonomous underwater vehicle}
\newacronym{colav}{COLAV}{collision avoidance}
\newacronym{colregs}{COLREGs}{International Regulations for Preventing Collisions at Sea}
\newacronym{cpa}{CPA}{closest point of approach}
\newacronym{ais}{AIS}{automatic identification system}
\newacronym{asv}{ASV}{autonomous surface vehicle}
\newacronym{imo}{IMO}{International Maritime Organization}
\newacronym{dof}{DOF}{{ degrees of freedom}}
\newacronym{pi}{PI}{proportional-integral}
\newacronym{fbcon}{FB}{feedback}
\newacronym{ffcon}{FF}{feedforward}
\newacronym{fffbcon}{FF-FB}{feedforward feedback}
\newacronym{fblcon}{FBL}{feedback-linearizing}
\newacronym{fffbccon}{FF-FB-C}{feedforward-feedback course}
\newacronym{dw}{DW}{dynamic window}
\newacronym{enc}{ENC}{electronic nautical charts}
\newacronym{bc-mpc}{BC-MPC}{branching-course model predictive control}
\newacronym{sb-mpc}{SB-MPC}{simulation-based model predictive control}
\newacronym{mpc}{MPC}{model predictive control}
\newacronym{nlp}{NLP}{nonlinear program}
\newacronym{spns}{SPNS}{single point neighbor search}
\newacronym{ncdm}{NCDM}{neighbor course distribution method}
\newacronym{pdf}{PDF}{probability density function}
\newacronym{gmm}{GMM}{Gaussian mixture model}
\newacronym{cvm}{CVM}{constant velocity model}
\newacronym{arpa}{ARPA}{automatic radar plotting aid}
\newacronym{solas}{SOLAS}{Safety of Life at Sea}
\newacronym{ocp}{OCP}{optimal control problem}

\newacronym{vo}{VO}{velocity obstacles}
\newacronym{rra*}{R-RA*}{rule-based repairing A*}
\newacronym{rrt}{RRT}{rapidly exploring random tree}
\newacronym{sog}{SOG}{speed over ground}
\newacronym{rot}{yaw rate}{rate of turn}
\newacronym{mr}{MR}{Maritime Robotics}
\newacronym{pdaf}{PDAF}{probabilistic data association filter}
\newacronym{ros}{ROS}{Robot Operating System}
\newacronym{los}{LOS}{line-of-sight}
\newacronym{ned}{NED}{North-East-Down}
\newacronym{osd1}{OSD1}{Ocean Space Drone 1}

\makeglossaries

\def\keyFont{\fontsize{8}{11}\helveticabold }
\def\firstAuthorLast{Eriksen {et~al.}} 
\def\Authors{Bjørn-Olav H.\ Eriksen\,$^{*}$, Glenn Bitar, Morten Breivik and Anastasios M.\ Lekkas}
\def\Address{Centre for Autonomous Marine Operations and Systems, Department of Engineering Cybernetics, Norwegian University of Science and Technology (NTNU), O. S. Bragstads Plass 2D, 7491 Trondheim, Norway}
\def\corrAuthor{Bjørn-Olav H.\ Eriksen}

\def\corrEmail{bjorn-olav.h.eriksen@ieee.org}

\makeatletter
\newcommand{\customlabel}[2]{%
   \protected@write \@auxout {}{\string \newlabel {#1}{{#2}{\thepage}{#2}{#1}{}} }%
   \hypertarget{#1}{}
}
\makeatother
\newcommand{\subfigCaption}[3]{\begin{tabulary}{.9\linewidth}{lL} \textbf{(#1)} & #3\end{tabulary}\customlabel{#2}{Figure \number\numexpr\value{figure}+1\relax#1}}

\newcommand{\astar}{\texorpdfstring{A$^\star$}{A*}}

\newcommand{\bs}{\boldsymbol}
\newcommand{\tr}{^\top}

\newcommand{\vect}[1]{\bs{#1}}
\newcommand{\real}{\mathbb{R}}

\newcommand{\nat}{\mathbb{N}}

\DeclareMathOperator{\scirc}{S}

\DeclarePairedDelimiter\abs{\lvert}{\rvert}%
\DeclarePairedDelimiter\norm{\lVert}{\rVert}%


\newcounter{step}

\begin{document}
\onecolumn
\firstpage{1}

\title[Hybrid COLAV for ASVs]{Hybrid Collision Avoidance for ASVs Compliant with COLREGs Rules 8 and 13--17} 

\author[\firstAuthorLast ]{\Authors} 
\address{} 
\correspondance{} 

\extraAuth{}

\maketitle

\begin{abstract}

\section{}
This paper presents a three-layered hybrid \gls{colav} system for \acrlongpl{asv}, compliant with rules 8 and 13--17 of the \gls{colregs}.
The \gls{colav} system consists of a high-level planner producing an energy-optimized trajectory, a \acrlong{mpc} based mid-level \gls{colav} algorithm considering moving obstacles and the \gls{colregs}, and the \acrlong{bc-mpc} algorithm for short-term \gls{colav} handling emergency situations in accordance with the \gls{colregs}.
Previously developed algorithms by the authors are used for the high-level planner and short-term \gls{colav}, while we in this paper further develop the mid-level algorithm to make it comply with \gls{colregs} rules 13--17.
This includes developing a state machine for classifying obstacle vessels using a combination of the geometrical situation, the distance and time to the \gls{cpa} and a new \gls{cpa}-like measure.
The performance of the hybrid \gls{colav} system is tested through numerical simulations for three scenarios representing a range of different challenges, including multi-obstacle situations with multiple simultaneously active \gls{colregs} rules, and also obstacles ignoring the \gls{colregs}.
The \gls{colav} system avoids collision in all the scenarios, and follows the energy-optimized trajectory when the obstacles do not interfere with it.
\tiny
 \keyFont{ \section{Keywords:} Hybrid collision avoidance, Autonomous surface vehicle (ASV), COLREGs, COLREGs compliant, Model predictive control (MPC), Energy-optimized control} 
\end{abstract}
\glsresetall

\section{Introduction}

Motivated by the potential for reduced costs and increased safety, the maritime industry is rapidly moving towards autonomous operations.
Following groundbreaking advances in the automotive industry, many sectors within the maritime industry are considering the benefits of autonomy, which includes more environmentally friendly operations.
For instance, the agricultural chemical company \emph{Yara} together with the maritime technology supplier \emph{Kongsberg Maritime} are developing the electrical autonomous container vessel \emph{Yara Birkeland}, which aims to replace 40 thousand yearly truck journeys in urban eastern Norway\footnote{\url{https://www.wsj.com/articles/norway-takes-lead-in-race-to-build-autonomous-cargo-ships-1500721202}, Accessed 2019-05-22.}.
Another example is the world's first autonomous car ferry, \emph{Falco}, developed by \emph{Rolls-Royce} (recently bought by Kongsberg Maritime) and \emph{Finferries}.
In 2018, \emph{Falco} navigated autonomously between two ports in Finland\footnote{\url{https://www.marinemec.com/news/view,rollsroyce-and-finferries-demonstrate-worlds-first-fully-autonomous-ferry_56102.htm}, Accessed 2019-04-11.}.
Reports state that in excess of \SI{75}{\percent} of maritime accidents are due to human errors \citep{Chauvin2011,Levander2017}, indicating that there is also a potential for increased safety in addition to the economical and environmental benefits.

An obvious prerequisite for autonomous ship operations is the development of robust and well-functioning \gls{colav} systems.
In addition to generating collision-free maneuvers, a \gls{colav} system must adhere to the ``rules of the road'' of the oceans, i.e. the \gls{colregs} \citep{Cockcroft2004}.
These rules are written for human ship operators and include qualitative requirements on how to perform safe and readily observable maneuvers.
Part B of the \gls{colregs} concern steering and sailing, and includes the following rules that are the most relevant to a motion control system:
\begin{desclist}{\hspace{.3cm}\bf}{ }[Rules 13--15]
	\item[Rule 8] Requires maneuvers to be readily observable and to be done in ample time.
	\item[Rules 13--15] Describe the maneuvers to perform in cases of overtaking, head-on and crossing situations.
	Participants in crossing situations are defined by the terms \emph{give-way} and \emph{stand-on} vessels.
	\item[Rule 16] Requires that a give-way vessel must take early and substantial action to keep clear of the stand-on vessel.
	\item[Rule 17] Consists of two main parts. 
	The first part requires a stand-on vessel to maintain its course and speed, while the second part allows/requires\footnote{Rule 17 \emph{allows} the stand-on vessel to maneuver when it becomes apparent that the give-way vessel does maneuver to avoid collision.
	If the vessels are so close that the give-way vessel cannot avoid collision by itself, Rule 17 \emph{requires} the stand-on vessel to maneuver.} a stand-on vessel to take action to avoid collision if the give-way vessel is not taking action.
\end{desclist}
Since the rules are written for humans, with few quantitative figures, a challenge for autonomous operation is to quantify them into behaviors that can be executed algorithmically.
The focus of the work in this paper is to do that, and to design a hybrid \gls{colav} system that performs motion planning and generates maneuvers in compliance to rules 8 and 13--17 of the \gls{colregs}.

A number of \gls{colav} approaches considering the \gls{colregs} have been proposed in the past.
This includes algorithms using \acrlong{sb-mpc} \citep{Hagen2018}, \acrlong{vo} \citep{Kuwata2014}, \acrlong{rra*} \citep{Campbell2014} and interval programming \citep{Benjamin2006}.
All these approaches are single-layer approaches, where one algorithm solves the complete \gls{colav} problem.

Another approach to the \gls{colav} problem is to use a hybrid architecture, where the task of planning an obstacle-free path or trajectory, complying with the \gls{colregs} and ultimately performing safe maneuvers is divided into layers in a control hierarchy.
The idea of hybrid architectures is to divide the subtasks of the \gls{colav} problem into multiple algorithms, exploiting their complementary strengths.
This also has the side effect of making it easier for human operators or supervisors to understand the system.
Most single-layer algorithms use sample-based approaches that consider a finite number of discrete control inputs, as opposed to conventional gradient-based search algorithms.
The reason for this is that many gradient-based algorithms are not sufficiently numerically robust, not allowing a \gls{colav} system to solely rely on such an algorithm.
This issue can be handled in hybrid architectures, constrained by the bottom-level algorithm being numerically robust and able to handle extraordinary situations where the other algorithms fail.
Hence, hybrid architectures also allows using gradient-based algorithms, which are able to solve problems with large search spaces more efficiently than sample-based algorithms.
The works by \citet{Svec2013} and \citet{Loe2008} are examples of two-layered hybrid \gls{colav} architectures.
The top layers perform trajectory planning among static obstacles, while the bottom layers perform moving obstacle avoidance in compliance with \gls{colregs} rules 13--16.
\citet{Casalino2009} presents a three-layered hybrid \gls{colav} system where the top layer also performs trajectory planning amongst static obstacles.
The middle layer avoids moving obstacles, while the bottom layer implements safety functions for handling cases where the two other layers fail.
This approach does, however, not consider the \gls{colregs}.

\begin{figure}[tb]
	\centering
	\includegraphics[width=.7\linewidth]{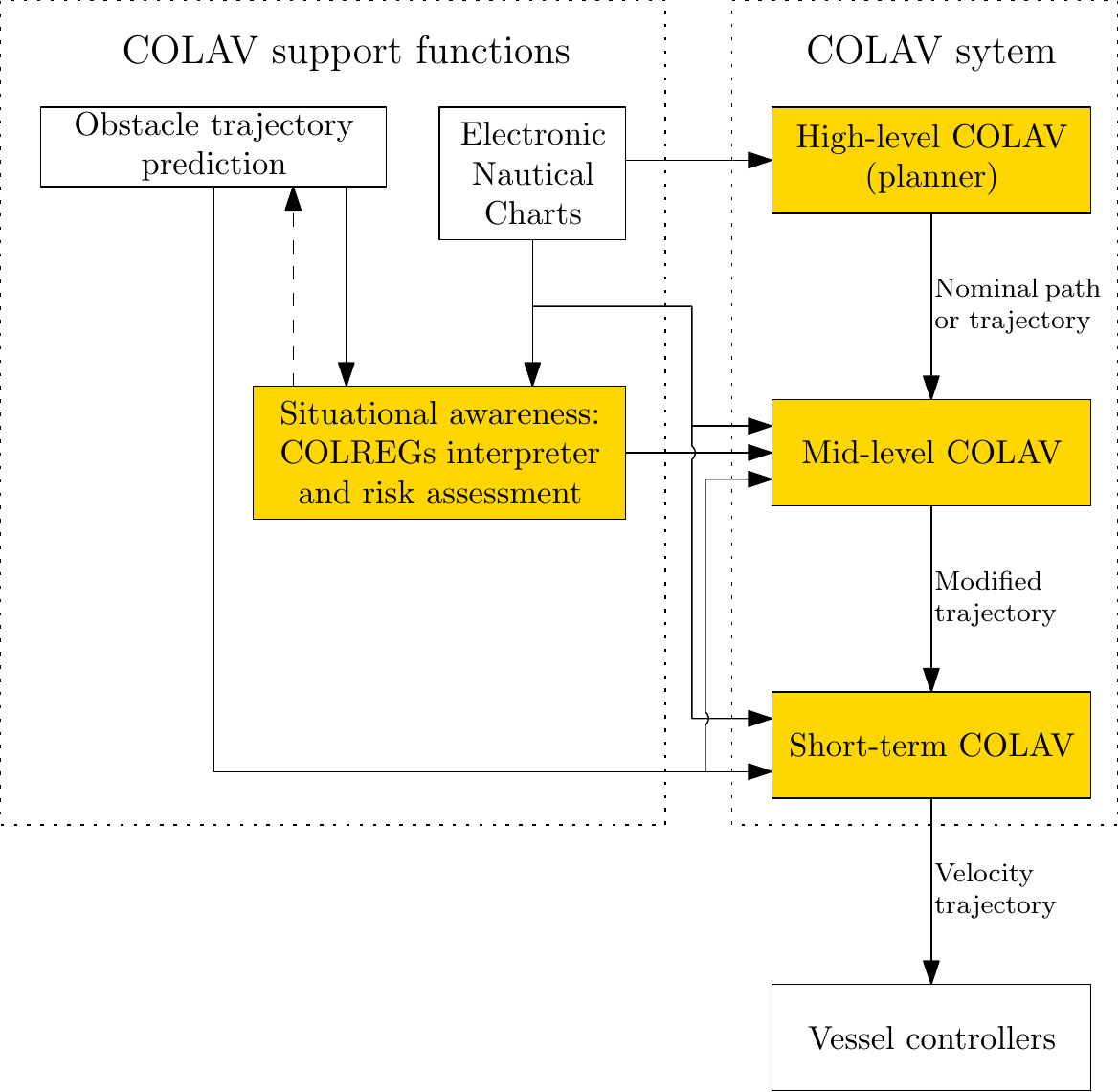}
	\caption{%
		Hybrid architecture with three \acrshort{colav} layers, where the highlighted functions mark the areas of interest in this article.
		The \acrshort{colav} system consists of a high-level planner, a mid-level \acrshort{colav} algorithm and a short-term \acrshort{colav} algorithm.
		The \acrshort{colav} system is supported by data from electronic nautical charts, represented in a suitable manner for the algorithms, as well as situational awareness functions that track and predict obstacles, interpret the \acrshort{colregs} and perform risk assessment.
	}
	\label{fig:hybrid-architecture}
\end{figure}
\autoref{fig:hybrid-architecture} shows a three-layered hybrid \gls{colav} system for an \gls{asv}.
The authors have previously worked extensively on different components of this architecture.
Examples include high-level \gls{colav} algorithms \citep{Bitar2018,Bitar2019cams}, a mid-level algorithm \citep{Eriksen2017b,Bitar2019ecc}, short-term algorithms \citep{Eriksen2018,Eriksen2019,Eriksen2019b} and the development of high-performance vessel controllers \citep{Eriksen2017,Eriksen2018b}.

In this paper, we demonstrate the three-layered hybrid \gls{colav} shown in \autoref{fig:hybrid-architecture} by combining and extending the \gls{colav} algorithms developed in \citep{Bitar2019ecc,Eriksen2017b,Bitar2019cams,Eriksen2019,Eriksen2019b}.
The high-level planner has a long temporal horizon, and finds an energy-optimized nominal trajectory from an initial to a goal position considering static obstacles.
Since the high-level planner only considers static information, it is intended to be run offline, but it can also be run online, for instance if new static obstacles are detected.
The mid-level algorithm attempts to follow this nominal trajectory, while performing \gls{colav} of moving obstacles in compliance with \gls{colregs} rules 8, 13--16 and the first part of Rule 17.
The mid-level algorithm is run periodically with a shorter temporal horizon than the high-level algorithm, and produces a modified trajectory which is passed to the short-term layer.
Both the high-level and mid-level algorithms use gradient-based optimization.
The short-term algorithm attempts to follow the modified trajectory, while it in compliance with the second part of Rule 17 handles situations where obstacles ignore the \gls{colregs}.
This algorithm also handles other emergency situations, and uses sample-based optimization to achieve a high level of robustness, ensuring safe operation if the mid-level algorithm fails to find a solution.
The following list summarizes our contributions:
\begin{itemize}
	\item The high-level planner from \citep{Bitar2019cams} is modified to include the mathematical model of the \emph{Telemetron} \gls{asv} in \citep{Bitar2019ecc}, including ocean currents.
	\item The development of a state-machine-based \gls{colregs} interpretation scheme.
	\item The mid-level \gls{colav} from \citep{Bitar2019ecc} is modified to include rules 13--16 and the first part of Rule 17.
	\item The \gls{bc-mpc} algorithm for short-term \gls{colav} is modified to reduce oscillatory behavior in turns.
	\item The three-layered \gls{colav} system is verified in simulations and shown to be compliant with rules 8 and 13--17.
\end{itemize}

The rest of the paper has the following structure:
The mathematical model of the \gls{asv} \emph{Telemetron} is described in \autoref{sec:modeling}.
The high-level planner, mid-level and short-term \gls{colav} algorithms are described in sections~\ref{sec:hi-level} to~\ref{sec:short-term}, respectively.
In \autoref{sec:results} we present and discuss the simulation scenarios and results, and we conclude the paper in \autoref{sec:conclusion}.

\section{ASV modeling}
	\label{sec:modeling}
\begin{figure}
	\centering
	\includegraphics[width=.8\linewidth]{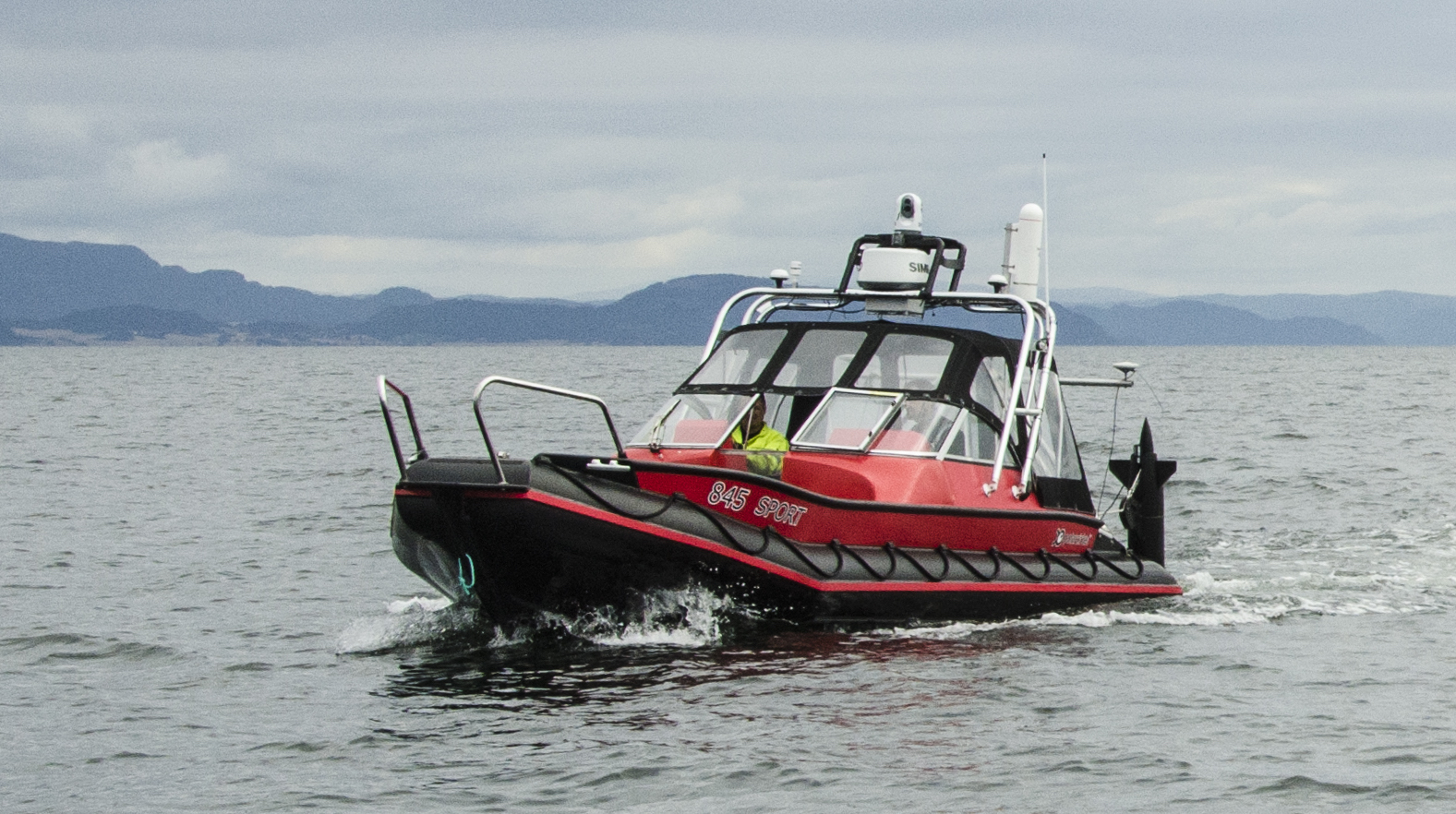}
	\caption{\label{fig:telemetron}The Telemetron \acrshort{asv}, designed for both manned and unmanned operations. Courtesy of Maritime Robotics.}
\end{figure}
The vessel of interest in this article is the Telemetron \gls{asv}, which is owned and operated by the Norwegian company \acrlong{mr} and shown in \autoref{fig:telemetron}.
The Telemetron \gls{asv} is a high-speed dual-use vessel propelled by a steerable outboard engine, capable of speeds up to $18~\si{\meter\per\second}$.

\citet{Eriksen2017} presents a model of the Telemetron \gls{asv}, which is extended to include ocean currents in \citep{Bitar2019ecc}.
The model has the form
\begin{equation}\label{eq:asvmodel}
	\begin{gathered}
		\dot{\bs\eta} = \bs R(\psi)\bs x_r + \begin{bmatrix} \bs V_c\tr & 0 \end{bmatrix}\tr \\
		\bs M(\bs x_r) \dot{\bs x}_r + \bs \sigma(\bs x_r) = \bs\tau\,,
	\end{gathered}
\end{equation}
where $\bs \eta = [x,y,\psi]\tr\in\real^2\times S$ is the vessel pose and $\bs V_c = [V_x,V_y]\tr$ describes the ocean current, both in the Earth-fixed \acrlong{ned} frame $\{n\}$.
The vector $\bs x_r = [u_r,r]\tr \in \mathbb{X}_r \subset \real^2$ is the vessel velocity under the assumption of zero relative sway motion \citep{Bitar2019ecc}, where the set $\mathbb{X}_r$ describes the vessel-feasible steady-state velocities where \eqref{eq:asvmodel} is valid.
The transformation matrix $\bs R(\psi)$ is given by the heading $\psi\in S$ as
\begin{equation}
	\bs R(\psi) =
	\begin{bmatrix}
		\cos \psi & 0 \\
		\sin \psi & 0 \\
		0 & 1
	\end{bmatrix}\,,
\end{equation}
while $r\in\real$ describes the vessel yaw-rate.
The matrix $\bs M(\bs x_r)$ is a state-dependent inertia matrix, while $\bs \sigma(\bs x_r)$ and $\bs \tau = [\tau_m,\tau_\delta]\tr\in\mathbb{U}\subset\real^2$ describe the vessel damping and control input, respectively.
The set $\mathbb{U}$ describes the control inputs where \eqref{eq:asvmodel} is valid.

\section{High-level planner}
	\label{sec:hi-level}

To plan the \gls{asv}'s nominal trajectory, we use a high-level trajectory planner developed in \citep{Bitar2019cams}.
This trajectory planner uses the \gls{asv} model described in \autoref{sec:modeling} to generate an energy-optimized trajectory between the start and goal positions.
The planning algorithm combines an \astar{} implementation and an \gls{ocp} solver to generate a feasible and optimized trajectory.

The high-level planning algorithm consists of three steps:
First the \astar{} implementation finds the shortest piecewise linear path between the start and goal position.
Secondly, artificial temporal information is added to the path, converting it to a trajectory of states and inputs.
Finally, the trajectory is used as an initial guess for an \gls{ocp} solver, which finds a locally energy-optimized trajectory near the shortest path.
All steps account for static obstacles in the form of elliptical boundaries.

\subsection{Static obstacles} 
\label{sub:static_obstacles}

The elliptical boundaries are described with the inequality:
\begin{equation}\label{eq:ellipsis_simple}
	\left(\frac{x-x_c}{x_a}\right)^2 + \left(\frac{y-y_c}{y_a}\right)^2 \geq 1\,,
\end{equation}
where $x_c$ and $y_c$ is the ellipsis center, and $x_a,y_a>0$ are the ellipsis major and minor axes, respectively.
To allow for angled obstacles, the ellipses are rotated clockwise by an angle $\alpha$.
We add a small constant $\epsilon > 0$ to each side of the inequality, and take the logarithm to arrive at the following obstacle representation:
\begin{multline}
	\label{eq:static-ellipse}
		h_{o}(x, y, x_c, y_c, x_a, y_a, \alpha) = - \log \bigg[\left( \frac{ (x - x_c) \cos \alpha + (y - y_c) \sin \alpha}{x_a} \right)^2 \\
		+ \left( \frac{-(x - x_c) \sin \alpha + (y - y_c) \cos \alpha}{y_a} \right)^2 + \epsilon \bigg] + \log(1 + \epsilon) \leq 0\,.
\end{multline}
The logarithm operation is applied to reduce the numerical range of the inequality, which helps with numerical stability of the subsequently described solver, and the constant $\epsilon$ is included to avoid singularities when \eqref{eq:static-ellipse} is evaluated for $(x,y) \to (x_c,y_c)$ \citep{Bitar2019ecc}.


\subsection{Trajectory generation and optimization} 
\label{sub:trajectory_generation_and_optimization}

From a scenario consisting of static obstacles, as mentioned in \autoref{sub:static_obstacles}, we find the piecewise linear shortest path by performing an \astar{} search on a uniformly decomposed grid.
The resulting path is converted to a time-parametrized full-state trajectory by assuming a constant forward velocity, and connecting the shortest path with straight segments and circle arcs.
The constant forward velocity is
\begin{equation}
	u_{\text{nom}} = \frac{L_{\text{path}}}{t_{\text{max}}}\,,
\end{equation}
where $L_{\text{path}}$ is the length of the connected path, and $t_{\text{max}}$ is the maximum allowed time to complete the trajectory.
This full-state trajectory is then used as an initial guess to solve the \gls{ocp} that gives the energy-optimized trajectory:
\begin{subequations}
\label{eq:hi-ocp}
\begin{align}
	\label{eq:hi-ocp-cost}
	&\min_{\vect{z}(\cdot), \vect{\tau}(\cdot)} \int_{0}^{t_{\text{max}}} F_{\text{hi}}(\vect{z}(t), \vect{\tau}(t)) \mathrm d t \\
	\nonumber
	&\text{subject to} \\
	\label{eq:hi-ocp-dynamics}
	&\dot{\vect{z}}(t) = \vect{f}(\vect{z}(t), \vect{\tau}(t)) ~ \forall t \in [0, t_{\text{max}}] \\
	\label{eq:hi-ocp-ineq}
	&\vect{h}_{\text{hi}}(\vect{z}(t), \vect{\tau}(t)) \leq \vect{0} ~ \forall t \in [0, t_{\text{max}}] \\
	\label{eq:hi-ocp-boundary}
	&\vect{e}_{\text{hi}}(\vect{z}(0), \vect{z}(t_{\text{max}})) = \vect{0}\,.
\end{align}
\end{subequations}
The solution of this \gls{ocp} is a trajectory of states $\vect{z}(\cdot)$ and inputs $\vect{\tau}(\cdot)$ that minimizes the cost functional in \eqref{eq:hi-ocp-cost}.
The \gls{asv} model from \autoref{sec:modeling} is rewritten as $\dot{\vect{z}} = \vect{f}(\vect{z}, \vect{\tau})$, where $\vect{z} = [\vect{\eta}\tr, \vect{x}_r\tr]\tr$ and $\vect{f}(\vect{z}, \vect{\tau})$ represents \eqref{eq:asvmodel}.

The cost functional \eqref{eq:hi-ocp-cost} is chosen to minimize energy.
The cost-to-go function is
\begin{equation}
	F_{\text{hi}}(\vect{z}, \vect{\tau}) = K_e F_e(\vect{z}, \vect{\tau}) + K_\delta \tau_\delta^2,
\end{equation}
with tuning parameters $K_e, K_\delta > 0$.
The first term consists of a function that is proportional to mechanical work performed by the \gls{asv}:
\begin{equation}
	\label{eq:hi-cost-to-go}
	F_e(\vect{z}, \vect{\tau}) = \abs{\underbrace{n(\tau_m)^2 \cdot \cos \delta(\tau_\delta)}_{\text{$\propto$ surge force}} \cdot u_r} + \abs{\underbrace{n(\tau_m)^2 \cdot \sin \delta(\tau_\delta) \cdot L_m}_{\text{$\propto$ yaw moment}} \cdot r}\,.
\end{equation}
The function $n: \real^+ \to \real^+$ maps the control input $\tau_m$ to propeller angular velocity.
The function $\delta: \real \to \scirc$ maps the control input $\tau_\delta$ to outboard motor angle.
The second term in \eqref{eq:hi-cost-to-go} is a quadratic cost to yaw control, included to avoid issues with singularity when solving the \gls{ocp}.

The inequality constraints \eqref{eq:hi-ocp-ineq} observe state boundaries as well as the static obstacles as represented in \autoref{sub:static_obstacles}.
The boundary conditions \eqref{eq:hi-ocp-boundary} denote initial and final constraints, i.e.\ start and end states.

A detailed description of the transcription of the \gls{ocp} \eqref{eq:hi-ocp} to a \gls{nlp} using multiple shooting with $N_{\text{hi}}$ shooting intervals is found in \citep{Bitar2019cams}.


\section{Mid-level COLAV}
	\label{sec:mid-level}
The mid-level algorithm, initially presented in \citep{Eriksen2017b} and further developed in \citep{Bitar2019ecc}, is a \gls{mpc}-based algorithm intended for long-term \gls{colav}.
The algorithm utilizes gradient-based optimization, and takes both static and moving obstacles into account while attempting to follow an energy-optimized nominal trajectory from the high-level planner.
The algorithm produces maneuvers complying with Rule 8 of the \gls{colregs}, which requires maneuvers to be made in ample time and be readily observable for other vessels.
The optimization problem is formulated as a \gls{nlp}, which gives flexibility in designing the optimization problem.

In this section, the algorithm is extended to also consider \gls{colregs} rules 13--16 and the first part of Rule 17.

\subsection{The International Regulations for Preventing Collisions at Sea (COLREGs)}
The \gls{colregs} consists of a total of 37 rules and is divided into five parts \citep{Cockcroft2004}, where part B (rules 4--19) contains relevant rules on the conduct of vessels in proximity of each other.
The most relevant rules for designing \gls{colav} systems in part B are rules 8 and 13--17:
\begin{desclist}{\hspace{.3cm}\bf}{ }[Rule 17]
	\item[Rule 8] \textbf{Action to avoid collision.}
	This rule states that actions taken to avoid collision should be large enough to be readily observable of other ships, implying that series of small alternations in speed and/or course should not be applied.
	The rule also recommends that course changes should be prioritized over speed changes if there is enough free space available, and that maneuvers must be made in ample time.
	\item[Rule 13] \textbf{Overtaking.}
	This rule states that a vessel is overtaking another if it approaches the other vessel with a course more than $22.5\si{\degree}$ abaft her beam.
	The overtaking vessel has to stay clear of the overtaken vessel, but there is no statements on which side of the vessel one should pass.
	\item[Rule 14] \textbf{Head on.}
	When two power-driven vessels approach each other on reciprocal, or nearly reciprocal, courses, they are in a head-on situation.
	In such a situation, both vessels should change their course to starboard, passing each other port-to-port, as shown in \ref{subfig:head_on_colregs}.
	This rule states no explicit definition on what should be considered to be reciprocal, or nearly reciprocal, courses, but court decisions indicate head-on situations exist for opposing courses $\pm6\si{\degree}$.
	Notice that the rule does not include sailing vessels, which are covered by Rule 12.
	\item[Rule 15] \textbf{Crossing.}
	When two vessels approach each other such that the situation is not a head on or an overtaking, it is a crossing situation.
	The vessel with the other one to her starboard side is deemed the give-way vessel, while the other vessel is deemed the stand-on vessel.
	As shown in \ref{subfig:crossing_colregs}, the give-way vessel should maneuver to avoid collision, preferably by passing behind the stand-on vessel, while the stand-on vessel should keep her speed and course.
	\item[Rule 16] \textbf{Action by the give-way vessel.}
	Every vessel which is required to keep out of the way of another vessel should take early and large enough action to safely avoid collision.
	\item[Rule 17] \textbf{Action by the stand-on vessel.}
	This rule requires that a stand-on vessel should keep its current speed and course.
	The stand-on vessel may, however, maneuver to avoid collision if it becomes apparent that the give-way vessel is not taking appropriate actions to avoid collision.
	Furthermore, if the stand-on vessel finds itself so close to the obstacle that collision can not be avoided by the give-way vessel alone, the stand-on vessel should take such action which best aids to avoid collision.
	In a crossing situation, the stand-on vessel should avoid maneuvering to port, since this could lead to a collision if the give-way vessel maneuvers to starboard.
\end{desclist}

In the hybrid architecture illustrated in \autoref{fig:hybrid-architecture}, the mid-level algorithm is given the task of strictly enforcing \gls{colregs} rules 13--16 and the stand-on requirement of Rule 17, while also complying with Rule 8.
\begin{figure}
  	\centering
  	\begin{tabular}[b]{@{}cc@{}}
      	\includegraphics[width=.45\linewidth]{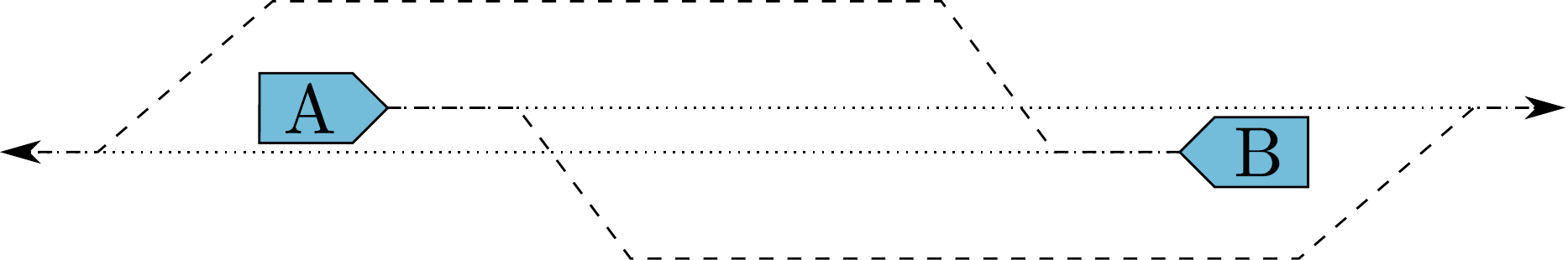} &
      	\includegraphics[width=.25\linewidth]{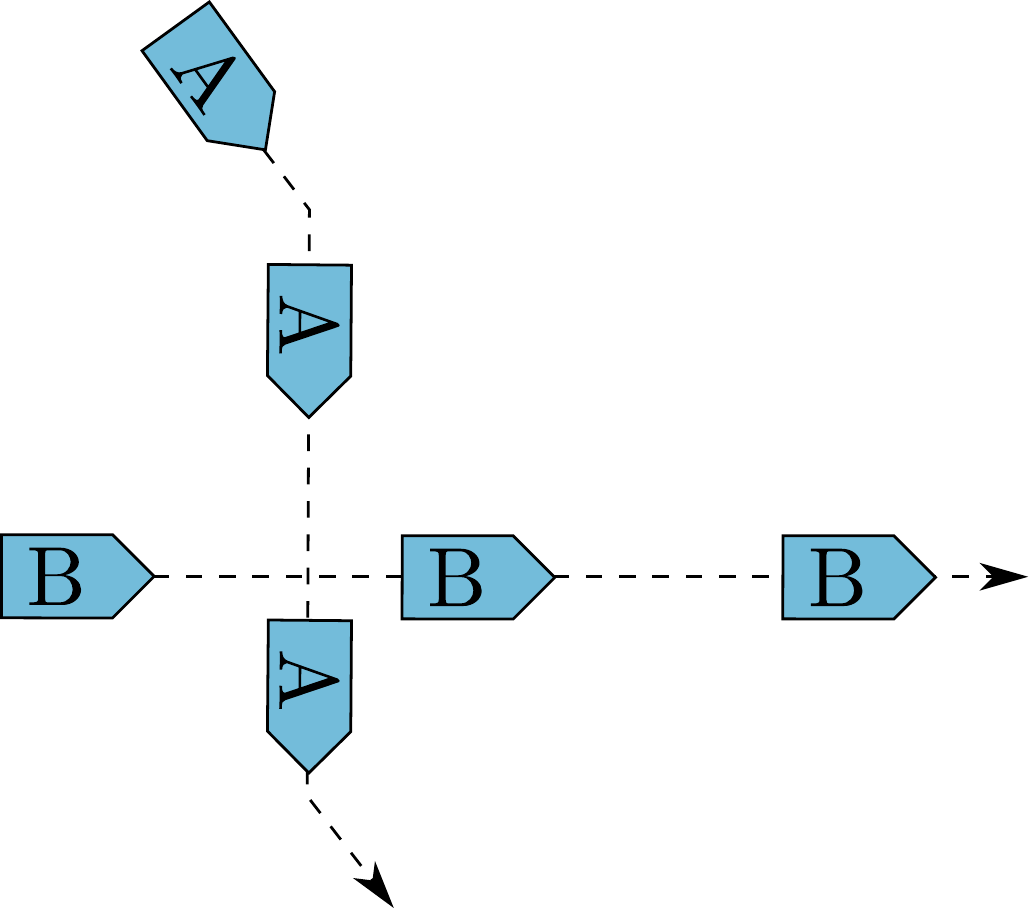} \\
      	\subfigCaption{A}{subfig:head_on_colregs}{Head-on situation.}&
      	\subfigCaption{B}{subfig:crossing_colregs}{Crossing situation.}
  	\end{tabular}
    \caption{\label{fig:colregs_situations}Illustration of head-on (A) and crossing (B) situations, and how they should be resolved.}
\end{figure}
In addition, we want the mid-level algorithm to comply with the first part of Rule 17, by not maneuvering to avoid collision in crossing situations if the ownship is the stand-on vessel.
The \gls{colav} system is inherently capable of adhering to the remaining requirement of Rule 17, where the stand-on vessel is allowed or required to maneuver, by having different prediction horizons and safety margins in the mid-level and short-term layers.
The \gls{bc-mpc} algorithm does not have any limitations of not maneuvering in stand-on situations, and will hence maneuver in stand-on situations if we come sufficiently close to the obstacle.

The mid-level algorithm as presented in \citep{Bitar2019ecc} only complies with Rule 8.
Further in this section, we therefore present improvements to the mid-level algorithm to make it comply with rules 13--16 and the stand-on requirement of Rule 17.

\subsection{COLREGs interpretation}
A commonly used concept for interpreting obstacles in \gls{colav} algorithms is to use assign a spatial region to obstacles, which the ownship should not enter.
This approach is commonly referred to as a \emph{domain-based} approach.
Specially designed ship domains are commonly used for interpreting the \gls{colregs} in \gls{colav} algorithms, where one usually require a larger clearance to obstacles if choosing maneuvers that violate the \gls{colregs} \citep{Eriksen2019,Szlapczynski2017}.
This approach is attractive since it continuously captures multiple \gls{colregs} rules, and does not require logic or discrete decisions.
However, such an approach does not strictly enforce the \gls{colregs} rules, since it will allow maneuvers violating the rules if they are large enough.
In addition, a ship-domain approach will not be able to strictly enforce the stand-on requirement of Rule 17, since a domain-based approach will avoid collision with all obstacles.
One could ignore obstacles with give-way obligations, but this would require an explicit \gls{colregs} interpretation which conflicts with domain-based approaches' core idea of implicit \gls{colregs} interpretation.
Therefore, we pursue an alternative approach to handling the \gls{colregs} in the mid-level algorithm.

To simplify the \gls{colregs} interpretation task, we look at the situation from a static perspective, assuming that the current \gls{colregs} situations are valid throughout the entire prediction horizon of the mid-level algorithm.
In reality, the \gls{colregs} situations may, however, change during the prediction horizon depending on both the ownship's and obstacles future trajectory.
For instance, an obstacle approaching from head on, but far enough away to not be considered as a danger may be put in a safe state.
Hence, the mid-level algorithm will (for the current iteration) act like no \gls{colregs} rule applies to this vessel for the entire prediction horizon, while the obstacle may get close enough during the prediction horizon to be considered as a head-on situation.
An \gls{mpc} scheme of only implementing a small part of the prediction horizon will reduce the implications of this, since the situation is reassessed each time mid-level algorithm is run, which justifies the assumption of considering the \gls{colregs} from a static perspective.
Investigating the possibilities for dynamically predicting future \gls{colregs} situations as part of the \gls{mpc} prediction will be considered as future work.

\subsubsection{State machine}
We propose to utilize a state machine in order to decide which \gls{colregs} rule is active with respect to each obstacle in the vicinity of the ownship.
The state machine contains the states:
\begin{desclist}{\hspace{.3cm}\bf}{ }[GW]
	\item[SF] Safe state. 
	This implies that the \gls{colregs} does not enforce any rule with respect to this obstacle.
	\item[OT] Overtaking state. 
	This implies that \gls{colregs} Rule 13 applies with respect to this obstacle. 
	The state machine does not discriminate on whether the ownship is overtaking another vessel or is being overtaken, but this can be done by looking at which vessel has the higher speed \citep{Tam2010}.
	\item[HO] Head-on state. 
	This implies that \gls{colregs} Rule 14 applies with respect to this obstacle.
	\item[GW] Give-way state. 
	This implies that \gls{colregs} Rule 15 applies with respect to this obstacle, and the ownship has to give way.
	\item[SO] Stand-on state. 
	This implies that \gls{colregs} Rule 15 applies with respect to this obstacle, and the ownship has to stand on.
	\item[EM] Emergency state. 
	This implies that the obstacle is so close and/or behaves unpredictably, such that special considerations must be made.
\end{desclist}

\begin{figure}[tb]
	\centering
	\includegraphics[width=.3\linewidth]{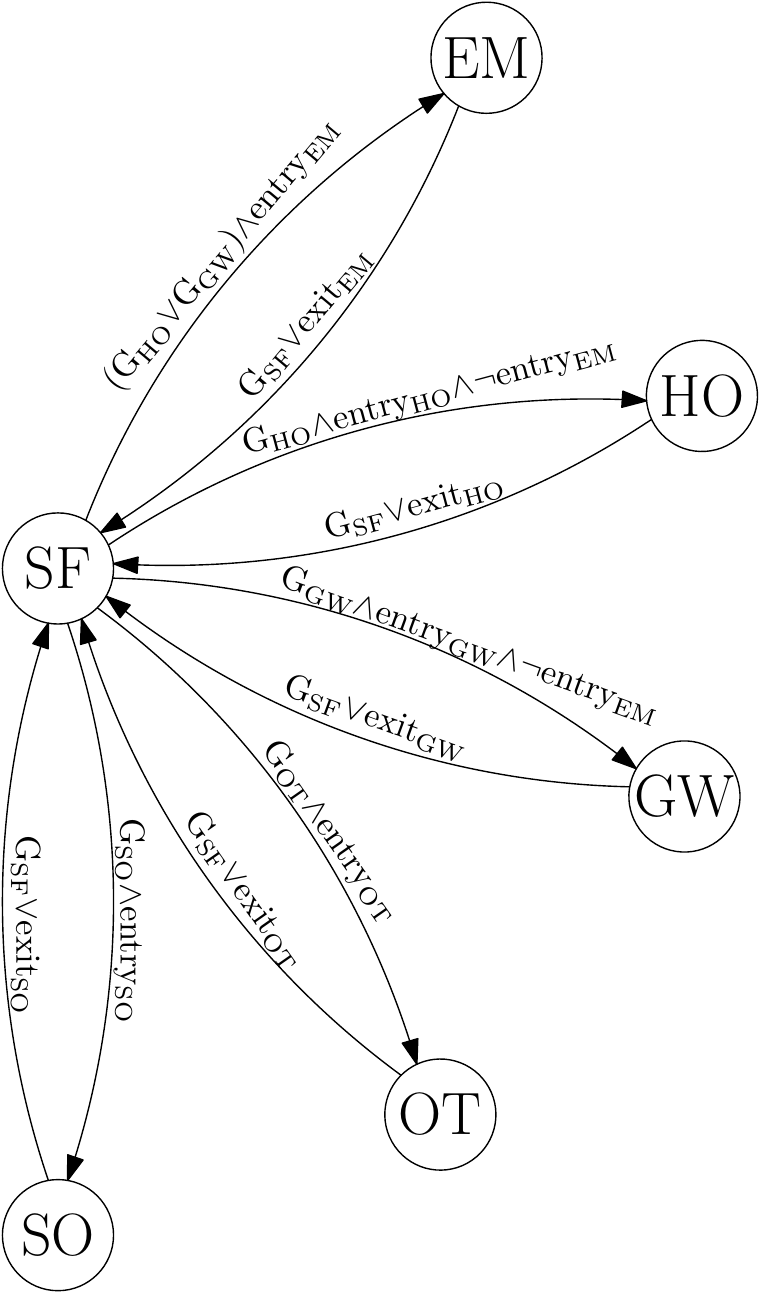}
	\caption{\label{fig:state-machine}\acrshort{colregs} state machine.
	The abbreviations ``G\textsubscript{SF}'', ``G\textsubscript{SO}'', ``G\textsubscript{OT}'', ``G\textsubscript{GW}'' and ``G\textsubscript{HO}'' denote geometrical situations, while ``entry\textsubscript{xx}'' and ``exit\textsubscript{xx}'' denote additional state-dependent entry and exit criterias.}
\end{figure}

As shown in \autoref{fig:state-machine}, all transitions have to go either from or to the safe state.
This implies that when the state machine decides that a \gls{colregs} (or emergency) situation exists with respect to an obstacle, it will not allow switching to another state without the situation being considered as safe first.
One could argue that it should be able to transition between specific states, like e.g. from head-on, give-way and overtaking to emergency.
This is an interesting topic, which should receive attention in the future.
To control the transitions between the different states, we combine the time to and distance at the \gls{cpa}, a \gls{cpa}-like measure of the time until a critical point and a geometrical interpretation of the situation.

\subsubsection{Entry and exit criteria}
\Gls{cpa} is a common concept in maritime risk assessment.
Given the current speed and course of the ownship and an obstacle, \gls{cpa} describes the time to the point where the two vessels are the closest, and the distance to the obstacle at this point.
Given the position and velocity vector of the ownship $\bs p, \bs v$ and an obstacle $\bs p_o, \bs v_o$, the time to \gls{cpa} is calculated as \citep{Kufoalor2018}
\begin{equation}
	t_{\text{CPA}} = \begin{cases}
		0 & \text{if }\norm{\bs v - \bs v_0}_2 \leq \epsilon \\
		\frac{(\bs p - \bs p_o) \cdot (\bs v - \bs v_o)}{\norm{\bs v - \bs v_o}^2_2} &\text{else},
	\end{cases}
\end{equation}
where $\epsilon > 0$ is a small constant in order to avoid division by zero in the case where the relative velocity between the ownship and obstacle is zero.
Given $t_{\text{CPA}}$, we calculate the distance between the vessels at \gls{cpa} as
\begin{equation}
	d_{\text{CPA}} = \norm{(\bs p + t_{\text{CPA}}\bs v) - (\bs p_o + t_{\text{CPA}}\bs v_o)}_2.
\end{equation}


While the \gls{cpa} is the point where the distance to an obstacle is at its minimum, the critical point is where the distance to an obstacle crosses underneath a critical distance $d_\text{crit}$.
This critical distance describes a minimum obstacle distance that the mid-level algorithm is designed for.
The time to the critical point $t_{\text{crit}}$ can be calculated by solving the equation
\begin{equation}
	\norm{(\bs p + t_{\text{crit}}\bs v) - (\bs p_o + t_{\text{crit}}\bs v_o)}_2 = d_{\text{crit}}\,.
\end{equation}
In the cases where the distance between the ships does not fall below $d_{\text{crit}}$, $t_{\text{crit}}$ is undefined.
Otherwise, there are generally two solutions.
The interesting solution is the one with the lowest $t_{\text{crit}}$ value, as this is when the obstacle enters the $d_{\text{crit}}$ boundary.

The state-machine entry criteria in \autoref{fig:state-machine} are defined as
\begin{equation}
	\begin{aligned}	
		\text{entry}_{i} &= \begin{cases}
			true & \text{if } d_{\text{CPA}} < \overline{d}_{\text{CPA}}^{i,\text{enter}} \land t_{\text{CPA}} \in[\underline{t}_{\text{CPA}}^{i,\text{enter}}, \overline{t}_{\text{CPA}}^{i,\text{enter}}] \\
			false & \text{otherwise}
		\end{cases}, \quad \forall i\in\{\text{SO,OT,GW,HO}\}\\
		\text{entry}_{\text{EM}} &= \begin{cases}
			true & \text{if } t_{\text{crit}} < \overline{t}_{\text{crit}}^{\text{EM,enter}} \land t_{\text{CPA}} > 0 \\
			false & \text{otherwise},
		\end{cases}
	\end{aligned}
\end{equation}
where $\overline{d}_{\text{CPA}}^{i,\text{enter}}, \underline{t}_{\text{CPA}}^{i,\text{enter}}$ and $\overline{t}_{\text{CPA}}^{i,\text{enter}}$ for $i\in\{\text{SO,OT,GW,HO}\}$ are tuning parameters denoting thresholds on $d_{\text{CPA}}$ and $t_{\text{CPA}}$ in order to satisfy the entry criteria for the stand-on, overtaking, give-way and head-on states. 
The tuning parameter $\overline{t}_{\text{crit}}^{\text{EM,enter}}$ denotes an upper limit on $t_{\text{crit}}$ in order to enter the emergency state.
The idea behind the stand-on, overtaking, give-way and head-on entry criterias are that in order for the obstacle to represent a risk, both $t_{\text{CPA}}$ and $d_{\text{CPA}}$ need to be within some tunable thresholds.
Situations with a very low $d_{\text{CPA}}$, but with a high $t_{\text{CPA}}$, will not trigger the entry criteria, since the situations will not occur in the near future.
Similarly, if $t_{\text{CPA}}$ is within the thresholds, but $d_{\text{CPA}}$ is large, this indicates a safe passing where risk of collision does not exist.
The lower bound on $t_{\text{CPA}}$ will typically be selected as zero, and is useful to distinguish between obstacles moving towards of away from the ownship.
For the emergency state, the entry criteria is based on the critical point, at which we are so close that the mid-level algorithm may struggle with providing meaningful maneuvers.
In addition to $t_{\text{crit}}$ being under the threshold $\overline{t}_{\text{crit}}^{\text{EM,enter}}$, we require that $t_{\text{CPA}}$ is positive, indicating that we are getting closer to the obstacle.
Currently, we only allow entering the emergency state if the situation is a geometrical give-way or head-on, since an overtaking situation represents a smaller danger and has less requirement for special consideration.

The state-machine exit criterias in \autoref{fig:state-machine} are defined as
\begin{equation}
	\begin{aligned}	
		\text{exit}_{i} &= \begin{cases}
			true & \text{if } d_{\text{CPA}} \geq \underline{d}_{\text{CPA}}^{i,\text{exit}} \lor t_{\text{CPA}} \notin[\underline{t}_{\text{CPA}}^{i,\text{exit}}, \overline{t}_{\text{CPA}}^{i,\text{exit}}] \\
			false & \text{otherwise}
		\end{cases}, \quad \forall i\in\{\text{SO,OT,GW,HO}\}\\
		\text{exit}_{\text{EM}} &= \begin{cases}
			true & \text{if } t_{\text{crit}} \geq \underline{t}_{\text{crit}}^{\text{EM,exit}} \lor t_{\text{CPA}} \leq 0 \\
			false & \text{otherwise},
		\end{cases}
	\end{aligned}
\end{equation}
where $\underline{d}_{\text{CPA}}^{i,\text{exit}}, \underline{t}_{\text{CPA}}^{i,\text{exit}}$ and $\overline{t}_{\text{CPA}}^{i,\text{exit}}$ for $i\in\{\text{SO,OT,GW,HO}\}$ are tuning parameters denoting thresholds on $d_{\text{CPA}}$ and $t_{\text{CPA}}$ in order to satisfy the exit criteria for the stand-on, overtaking, give-way and head-on states.
The exit criteria for the emergency state is satisfied if $t_{\text{crit}}$ is larger than the tuning parameter $\overline{t}_{\text{exit}}^{\text{EM,enter}}$, or $t_{\text{CPA}}$ is negative, implying that the obstacle is moving further away from the ownship.
Note that the exit criterias are obtained by negating the entry criterias, but with other thresholds in order to implement hysteresis to avoid shattering.
In general, we allow for different tuning parameters for the different states, but in our simulations we see that selecting the same tuning parameters for all states provides good results.
Therefore, we define:
\begin{equation}
	\begin{aligned}
		\overline{d}_{\text{CPA}}^{i,\text{enter}} &= \overline{d}_{\text{CPA}}^{\text{enter}} \\
		\underline{t}_{\text{CPA}}^{i,\text{enter}} &= \underline{t}_{\text{CPA}}^{\text{enter}} \\
		\overline{t}_{\text{CPA}}^{i,\text{enter}} &= \overline{t}_{\text{CPA}}^{\text{enter}},
	\end{aligned}
\end{equation}
and
\begin{equation}
	\begin{aligned}
		\underline{d}_{\text{CPA}}^{i,\text{exit}} &= \underline{d}_{\text{CPA}}^{\text{exit}} \\
		\underline{t}_{\text{CPA}}^{i,\text{exit}} &= \underline{t}_{\text{CPA}}^{\text{exit}} \\
		\overline{t}_{\text{CPA}}^{i,\text{exit}} &= \overline{t}_{\text{CPA}}^{\text{exit}}.
	\end{aligned}
\end{equation}

\subsubsection{Geometrical situation interpretation}
\begin{figure}
	\centering
	\includegraphics[width=.7\linewidth]{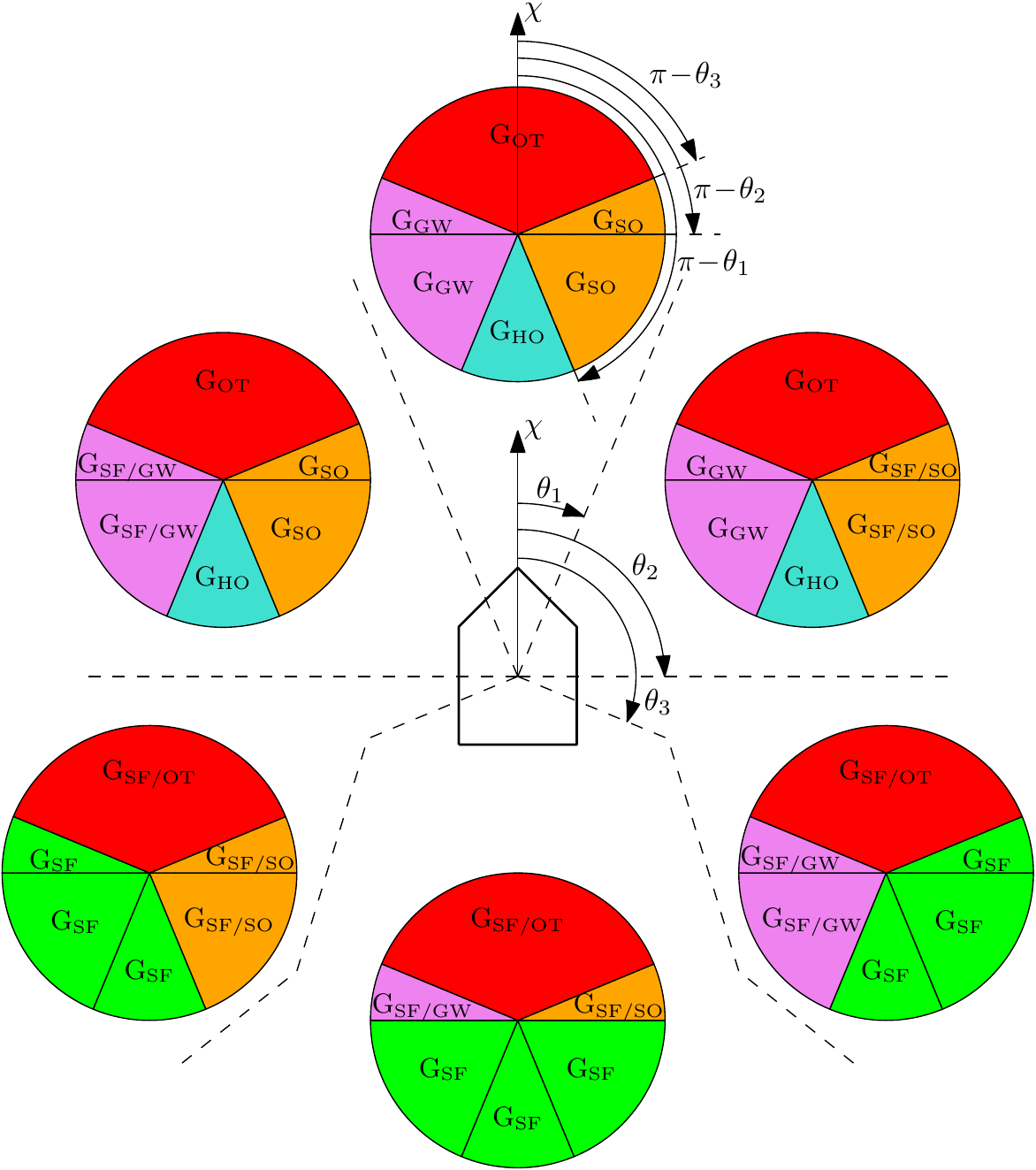}
	\caption{\label{fig:graphical-COLREGs-interpretation}Illustration of the geometrical \acrshort{colregs} interpretation, where the ownship course is denoted as $\chi$ and $\theta_1,\theta_2,\theta_3$ denote symmetrical regions given as $[22.5\si{\degree}, 90\si{\degree}, 112.5\si{\degree}]$.
	The circles illustrate obstacles in different relative bearing regions, and have a fixed orientation with respect to the ownship.
	The geometrical situations are color-coded and denoted as $\text{G}_i, i \in \{\text{SF, SO, OT, GW, HO}\}$ for safe, stand-on, overtaking, give-way and head-on situations, respectively.
	When two situations are given, like e.g. G\textsubscript{SF/SO}, we use the former (SF) if $t_\text{CPA}<0$ and the latter (SO) if $t_\text{CPA}\geq0$, analogous to the obstacle moving away or towards the ownship.
	To decide the geometrical situation, we first find which relative bearing region the obstacle resides in, before finding which obstacle region the obstacle's course resides in.
	The figure is inspired by \citep{Tam2010}.}
\end{figure}

\citet{Tam2010} present a geometrical interpretation scheme for deciding \gls{colregs} situations based on the relative position, bearing and course of the obstacle with respect to the ownship.
We base our geometrical interpretation on a slightly modified version of this scheme, where we include the sign of $t_{\text{CPA}}$ to distinguish between situations where the obstacle moves closer towards or farther away from the ownship.
The geometrical interpretation is shown in \autoref{fig:graphical-COLREGs-interpretation}, where the geometrical situation is obtained by finding which region the obstacle position and course resides in.
Notice that the head-on region is larger than the threshold of $\pm 6\si{\degree}$ as described by the \gls{colregs}.
The reason for this is that \citeauthor{Tam2010} recommend using a larger region of $22.5\si{\degree}$ in order to increase the robustness of the geometrical \gls{colregs} interpretation scheme.

\subsection{Interface to the high-level planner}
The high-level planner produces an energy-optimized nominal trajectory for the ownship to follow.
However, since the high-level planner does not consider moving obstacles, the speed is the only time-relevant factor of the desired trajectory.
In a case where the ownship for some reason, e.g. avoiding moving obstacles, lag behind the nominal trajectory, following the nominal trajectory in absolute time would cause a speed increase in order to catch up with it.
Therefore, the mid-level algorithm performs \emph{relative trajectory tracking}, where it tracks the nominal trajectory with a time offset $t_b \in \real$.
This results in a relative nominal trajectory for the mid-level algorithm:
\begin{equation}
	\bar{\bs p}_d(t) = \bs p_d(t+t_b),
\end{equation}
where $\bs p_d = [N_d(t), E_d(t) ]\tr$ is the nominal trajectory from the high-level planner.
The time offset $t_b$ is calculated each time the mid-level algorithm is run by solving a separate optimization problem, and is selected such that $\bar{\bs p}_d(t_0)$ is the point on the nominal trajectory closest to the ownship.
See \citep{Bitar2019ecc} for a detailed description of this concept.

\subsection{Optimization problem formulation}
The mid-level algorithm is formalized as an \gls{ocp}:
\begin{subequations}
\label{eq:ocp-mid}
\begin{align}
	\label{eq:ocp-mid-cost}
	&\min_{\vect{\eta}(\cdot),\vect{x}_r(\cdot)} ~ \phi(\vect{\eta}(\cdot), \vect{x}_r(\cdot)) \\
	\intertext{subject to}
	\label{eq:ocp-mid-dynamics}
	&\dot{\vect{\eta}}(t) = \bs R(\psi(t)) \vect{x}_r(t) +
	\begin{bmatrix}
	 	\vect{V}_c \\
	 	0
	 \end{bmatrix} ~ \forall t \in [t_0, t_0+T_h] \\
	\label{eq:ocp-mid-ineq}
	&\vect{h}_{\text{mid}}(\vect{\eta}(t), \vect{x}_r(t), t) \leq \vect{0} ~ \forall t \in [t_0, t_0+T_h] \\
	\label{eq:ocp-mid-boundary}
	&\vect{e}_{\text{mid}}(\vect{\eta}(t_0)) = \vect{0}\,,
\end{align}
\end{subequations}
where $T_h>0$ is the prediction horizon, $\phi(\cdot,\cdot)$ is the objective functional, \eqref{eq:ocp-mid-dynamics} contains a kinematic vessel model, \eqref{eq:ocp-mid-ineq} contains inequality constraints and \eqref{eq:ocp-mid-boundary} contains boundary constraints.

Analytical solutions of \glspl{ocp} are in general not possible to find.
A more common approach is to transcribe the \gls{ocp} to an \gls{nlp}, and solve that using a gradient optimization scheme.
In our case, we transcribe \eqref{eq:ocp-mid} into an \gls{nlp} with $N_p$ samples using multiple shooting, where the vessel model is discretized using 4th order Runge Kutta and the cost functional is discretized using forward Euler.
The resulting \gls{nlp} is given as
\begin{equation}
	\label{eq:NLP_mid}
	\begin{aligned}
		&\min_{\bs w, \bs \omega, \bs \mu, \bs \xi} ~ \phi_p(\bs{w}, \bs{\omega}, \bs{\mu}) + \phi_c(\bs{w}) + \phi_{\text{COLREGs}}(\bs{w}) + \phi_\xi(\bs \xi) \\
		&\text{subject to}\\
		&\bs{g}(\bs{w},\bs{\eta}(t_0)) = \bs{0} \\
		&\bs{h}(\bs{w},\bs \xi) \leq \bs{0}\\
		&\bar{\bs{h}}_k(\bs{\eta}_k, \bs{\omega}_k, \bs{\mu}_k, \bar{\bs{p}}_{d,k}) \leq \bs{0} \,\, \forall k\in\{1,\ldots,N_p\} \\
		& \bs \xi \geq \bs 0\,,
	\end{aligned}
\end{equation}
where $\vect{w} = [\vect{\eta}\tr_0, \vect{x}\tr_{r,0}, \ldots, \vect{\eta}\tr_{{N_p}-1}, \vect{x}\tr_{r,{N_p}-1}, \vect{\eta}\tr_{N_p}]\tr \in \real^{5N_p+3}$ is a vector of ${5N_p+3}$ decision variables and $\bar{\vect{p}}_{d,{1:N_p}} = [\bar{\vect{p}}_{d,1}, \bar{\vect{p}}_{d,2}, \ldots, \bar{\vect{p}}_{d,N_p}]$ is a sequence of desired positions. 
The vectors $\bs \omega \in \real^{2N_p}$, $\bs \mu \in \real^{2N_p}$ and $\bs \xi \in \real^{MN_p}$ contain slack variables, where $M$ is the number of moving obstacles to be included in the constraints.

The vector $\bs{g}(\bs{w},\bs{\eta}(t_0)) \in \real^{3N_p+3}$ contains shooting and boundary constraints, while $\bs{h}(\bs{w})\in \real^{(M+D+4)N_p}$, where $D$ is the number of static obstacles, contain inequality constraints ensuring \gls{colav} and steady-state vessel velocity feasibility.
The vectors $\bar{\bs{h}}_k(\bs{\eta}_k, \bs{\omega}_k, \bs{\mu}_k, \bar{\bs{p}}_{d,k}) \in \real^6$, $k\in\{1,N_p\}$ contain constraints on the slack variables $\bs \omega$ and $\bs \mu$.

In the following subsections, we describe the terms in \eqref{eq:NLP_mid} in more detail.

\subsubsection{Objective function}
The objective function contains four functions, where $\phi_p(\bs{w}, \bs{\omega}, \bs{\mu})$ introduces cost on deviating from the relative nominal trajectory $\bar{\bs p}_d(t)$, $\phi_c(\bs{w})$ introduces cost on using control input, $\phi_{\text{COLREGs}}(\bs w)$ is a \gls{colregs}-specific function and $\phi_\xi(\bs \xi)$ introduces slack variable cost.

To avoid that the \gls{nlp} changes behavior when moving away from the nominal trajectory, we wish to have linear growth in the position error function $\phi_p(\bs{w}, \bs{\omega}, \bs{\mu})$.
This is achieved by instead of using quadratic terms in the position error function, we use the Huber loss function which is quadratic around the origin and resembles the absolute value function above a given threshold $\sigma >0$:
\begin{equation}\label{eq:Huber}
	H(\rho) = \begin{cases} 
      \frac{1}{2}\rho^2 & \lvert \rho\rvert\leq \sigma \\
      \sigma(\lvert \rho \rvert - \frac{1}{2}\sigma) & \lvert \rho \rvert > \sigma\,.
   \end{cases}
\end{equation}
The Huber loss function has a discontinuous gradient, making it slightly complicated to implement in gradient-based optimization problems.
It can, however, be implemented in a continuous fashion by utilizing lifting, where slack variables are introduced to create a problem of a higher dimensionality which is easier to solve.
Using this technique, $\bar\phi_p(\vect{w}, \vect{\omega}, \vect{\mu})$ is defined as
\begin{equation}
	\bar\phi_p(\vect{w}, \vect{\omega}, \vect{\mu}) = K_p\sum_{k=1}^{N_p} \sigma \vect{1}\tr\vect{\omega}_{k} + \frac{1}{2}\vect{\mu}_{k}\tr\vect{\mu}_{k},
\end{equation}
where $K_p >0$ is a tuning parameter, and $\vect{\omega}_{k}\in\real^2$ and $\vect{\mu}_{k}\in\real^2$ are slack variables constrained by
\begin{equation}
	\bar{\vect{h}}_k(\vect{w}, \vect{\omega}, \vect{\mu}, \bar{\vect{p}}_{d,k}) = \begin{bmatrix} \vect{v_k} + \vect{\mu_k} + \vect{p}_k - \bar{\vect{p}}_{d,k} \\ \vect{v_k} + \vect{\mu_k} - (\vect{p}_k - \bar{\vect{p}}_{d,k}) \\ - \vect{\omega}_k \end{bmatrix} \leq \bs 0 \quad \forall k\in\{1, \ldots, N_p\},
\end{equation}
where $\bs p_k$ is the predicted vessel position at time step $k$, i.e. $\bs \eta_k = [\bs p_k\tr, \psi_k]\tr$.
See \citep{Bitar2019ecc} for more details.

Rule 8 of the \gls{colregs} requires that maneuvers are readily observable for other vessels, implying that speed and course changes should have a sufficiently large magnitude, and not be performed as a sequence of small changes.
In order to enforce this in the optimization problem, the control cost function $\phi_c(\bs{w})$ introduces a nonlinear cost on the change in speed and course, which makes the algorithm favor readily observable maneuvers.
The function is defined as
\begin{equation}\label{eq:ControlCost}
	\phi_c(\vect{w}) = \sum_{k=0}^{N_p-1} K_{\dot{U}} q_{\dot{U}}(\dot{U}_{k}) +K_{\dot \chi} q_{\dot{\chi}}(\dot \chi_{k}),
\end{equation}
where $K_{\dot{U}},K_{\dot{\chi}}>0$ are tuning parameters, while $q_{\dot{U}}(\dot{U}_{k})$ and $q_{\dot{\chi}}(\dot \chi_{k})$ are the nonlinear cost functions.
Notice that neither the \gls{sog} $U$ nor the course $\chi$ are elements of the search space, but they can be computed as $U = \sqrt{u^2 + v^2}$ and $\chi = \psi + \arcsin\frac{v}{U}$.
Their derivatives are then calculated by finite differencing.
See \citep{Eriksen2017,Bitar2019ecc} for more details on the control cost function.

The $\phi_{\text{COLREGs}}(\bs w)$ function introduces a \gls{colregs}-specific cost with respect to obstacles based on the rule currently applicable as defined by the state machine.
We hence tailor the \gls{nlp} to the current situation.
The function is defined as
\begin{multline}
	\phi_{\text{COLREGs}}(\bs w) = \sum_{k=1}^{N_p}\left[\sum_{i\in \mathcal{O}_{\text{HO}}}K_{\text{HO}}V_{\text{HO},i,k}(\bs p_k) + \sum_{i\in \mathcal{O}_{\text{GW}}}K_{\text{GW}}V_{\text{GW},i,k}(\bs p_k) \right.\\\left.+ \sum_{i\in \mathcal{O}_{\text{SO}}}K_{\text{SO}}V_{\text{SO},k}(\bs w) + \sum_{i\in \mathcal{O}_{\text{EM}}}K_{\text{EM}}V_{\text{EM},k}(\bs w)\right],
\end{multline}
where $\mathcal{O}_{\text{HO}}, \mathcal{O}_{\text{GW}},\mathcal{O}_{\text{SO}}$ and $\mathcal{O}_{\text{EM}}$ contain obstacles which are in the head-on, give-way, stand-on and emergency states, respectively, and $K_{\text{HO}},K_{\text{GW}},K_{\text{SO}},K_{\text{EM}}>0$ are tuning parameters.
The functions $V_{\text{HO},i,k}(\bs p_k), V_{\text{GW},i,k}(\bs p_k),V_{\text{SO},k}(\bs w)$ and $V_{\text{EM},k}(\bs w)$ describe functions capturing head-on, give-way, stand-on and emergency behavior with respect to obstacle $i$, respectively.
Notice that the head-on and give-way functions vary with both the obstacle number and time step number, which is due to the functions depending on the the given obstacles position and course at time step $k$.

For head-on situations, we define a potential function with a positive value on the obstacle's starboard side, and a negative value on its port side.
When used in the objective function, this will favor trajectories passing a head-on obstacle on its port side, in compliance with Rule 14 of the \gls{colregs}.
In addition, the potential function has an attenuation term, reducing the impact of the function when far away from an obstacle:
\begin{equation}
	V_{\text{HO},i,k}(\bs p) = \frac{\tanh\left(\alpha_{x,\text{HO}}(x_{0,\text{HO}} - x^{\{i,k\}})\right)}{2}\tanh(\alpha_{y,\text{HO}} y^{\{i,k\}}) \in (-1,1),
\end{equation}
where $\alpha_{x,\text{HO}},\alpha_{y,\text{HO}}>0$ are tuning parameters controlling the steepness of the head-on potential function and $\bar x_{0,\text{HO}} >0$ is a tuning parameter controlling the influence of the attenuating potential. 
The coordinate $(x^{\{i,k\}}, y^{\{i,k\}})$ is $\bs p$ given in obstacles \textit{i}'s course-fixed frame (in which the \textit{x}-axis points along the obstacle's course) at time step $k$, computed as
\begin{equation}
 	\begin{bmatrix} x^{\{i,k\}} \\ y^{\{i,k\}} \end{bmatrix} = \bs R(\chi_{i,k})\tr\left(\bs p - \bs p_{o,k,i}\right),
\end{equation}
where $\bs p_{o,k,i}$ and $\psi_{i,k}$ are the position and heading of obstacle $i$ at time step $k$.
The head-on potential function with parameters $\alpha_{x,\text{HO}} = \sfrac{1}{500}, \alpha_{y,\text{HO}} = \sfrac{1}{400}$ and $x_{0,\text{HO}}=1000~\si{\meter}$ is shown in \ref{subfig:head_on_potential}.
\begin{figure}
  	\centering
  	\begin{tabular}[b]{@{}cc@{}}
      	\includegraphics[width=.45\linewidth]{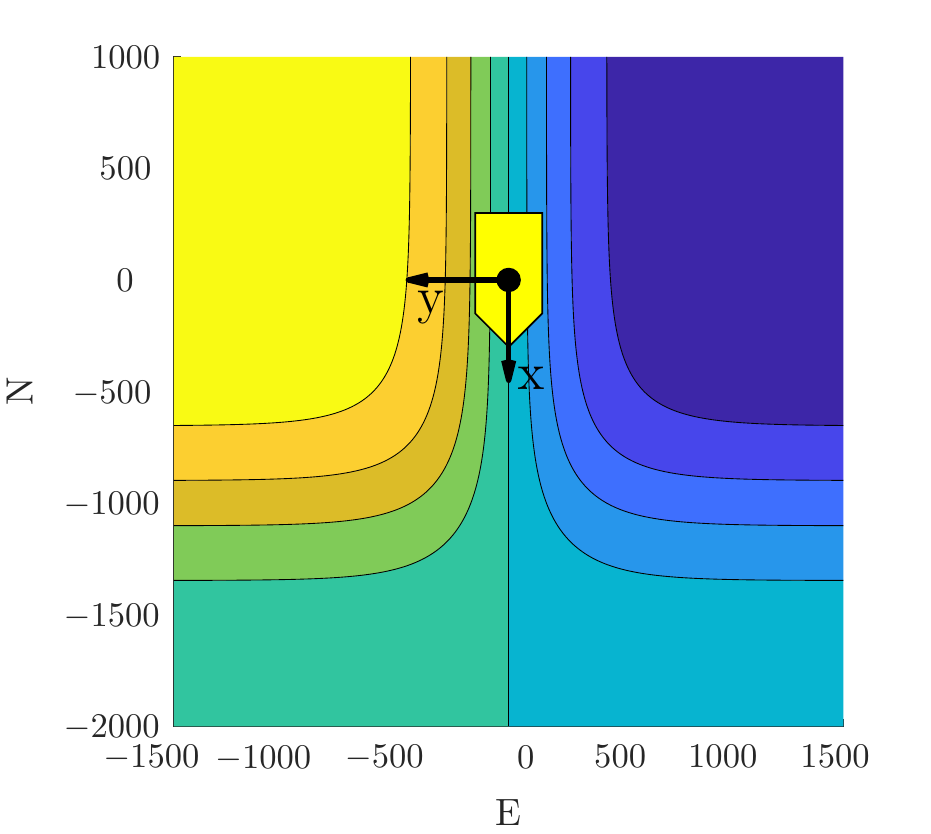} &
      	\includegraphics[width=.45\linewidth]{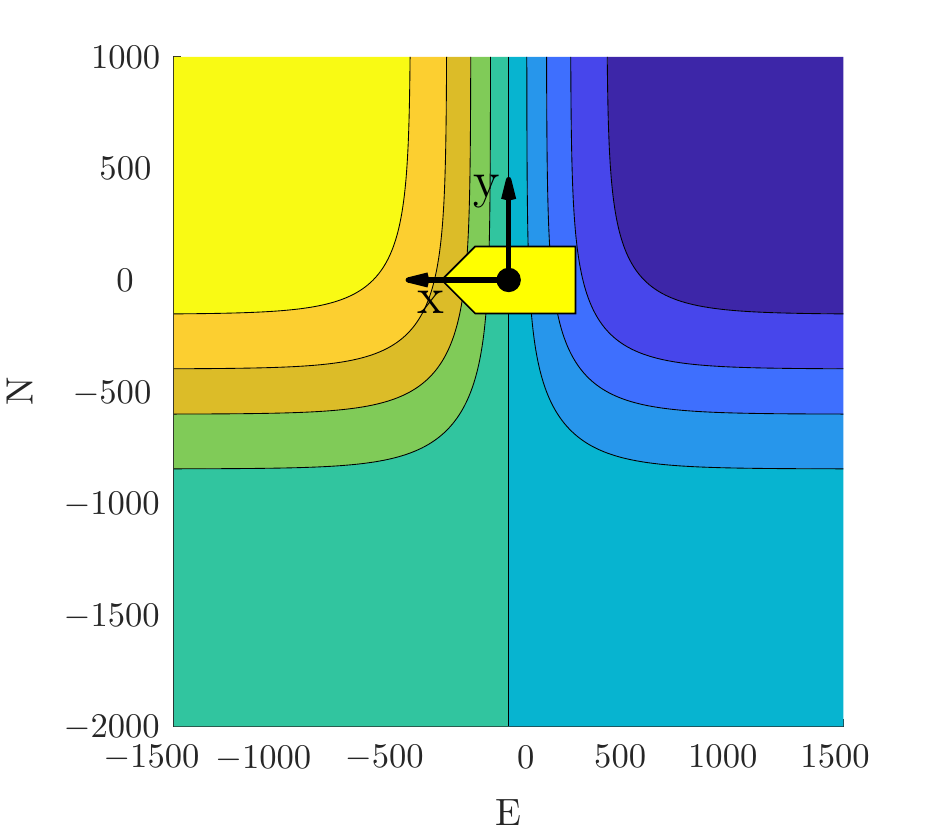} \\
      	\subfigCaption{A}{subfig:head_on_potential}{Head-on potential function.}&
      	\subfigCaption{B}{subfig:give_way_potential}{Give-way potential function.}
  	\end{tabular}
    \caption{\label{fig:potential_functions}Potential functions ensuring passing on the correct side in head-on and give-way situations.
    Yellow indicates a positive value, blue indicates a negative value, while the yellow patch and axis cross show the obstacle location and course-fixed coordinate system.
    Used in a minimization scheme, this will favor starboard maneuvers in head-on situations, and passing behind obstacles in give-way situations.
    Note that the obstacle here has zero sideslip, resulting in the heading and course pointing in the same direction.}
\end{figure}

For give-way situations, we define a similar potential function, but rotated such that the function is positive in front of an obstacle and negative behind it.
This will favor trajectories passing behind an obstacle, as desirable with respect to Rule 15 when a give-way obligation is active.
The give-way potential function is defined as
\begin{equation}
	V_{\text{GW},i,k}(\bs p) = \frac{\tanh\left(\alpha_{y,\text{GW}}(y^{i,k} - y_{0,\text{GW}})\right)}{2}\tanh(\alpha_{x,\text{GW}} x^{i,k}) \in (-1,1),
\end{equation}
where $\alpha_{x,\text{GW}}, \alpha_{y,\text{GW}} > 0$ control the steepness of the give-way potential function and $\bar y_{0,\text{GW}} < 0$ control the attenuation on the port side of an obstacle.
The give-way potential function with parameters $\alpha_{x,\text{GW}} = \sfrac{1}{400}, \alpha_{y,\text{GW}} = \sfrac{1}{500}$ and $y_{0,\text{GW}}=-500~\si{\meter}$ is shown in \ref{subfig:give_way_potential}.

In stand-on situations, we want the mid-level algorithm to disregard the obstacle and keep the current speed and course in order to comply with the first part of Rule 17.
One could simply constrain the algorithm to not maneuver, but this would be perilous in situations where the ownship simultaneously finds itself in a head-on or give-way situation. 
In such a situation it would be of extra importance to choose readily observable maneuvers, and we therefore design the stand-on cost with the same terms as used in the control cost \eqref{eq:ControlCost} to amplify the effect:
\begin{equation}
	V_{\text{SO},k}(\bs w) = K_{\dot{U}} q_{\dot{U}}(\dot{U}_{k}) +K_{\dot \chi} q_{\dot{\chi}}(\dot \chi_{k}).
\end{equation}

If an obstacle is in an emergency state, the obstacle is disregarded in the mid-level algorithm and left for the short-term algorithm to handle.
In such a situation, it is important that the mid-level algorithm behaves predictable, and we therefore use the same cost function as for stand-on situations:
\begin{equation}
	V_{\text{EM},k}(\bs w) = V_{\text{SO},k}(\bs w).
\end{equation}

The slack variable $\bs \xi$ is used in a homotopy scheme, which we introduce to avoid getting trapped in local minima around moving obstacles.
The homotopy scheme is described in further detail in \autoref{sec:obsHandling}.
The homotopy cost function $\phi_\xi(\bs \xi)$ introduces slack cost on $\bs \xi$:
\begin{equation}\label{eq:homotopy_cost}
	\phi_\xi(\bs \xi) = K_\xi \bs 1\tr \bs \xi,
\end{equation}
where $K_\xi > 0$ is iteratively increased as part of the homotopy scheme.

\subsection{Obstacle handling and steady-state feasibility}\label{sec:obsHandling}
The inequality constraint $\bs h(\bs w, \bs \xi) \leq \bs 0$ ensures \gls{colav} and steady-state feasibility with respect to actuator limitations.

Static obstacles are handled similarly as in the high-level algorithm, with \eqref{eq:static-ellipse} representing an elliptical obstacle with center $(x_c,y_c)$, angle $\alpha$ and major and minor axes $x_a$ and $y_a$, respectively.
The constraint \eqref{eq:static-ellipse} needs to be enforced at each time step.
Hence, for the $i$-th static obstacle, we define the constraint
\begin{equation}\label{eq:mid-level-static-obs}
	\vect{h}_{s_i}(\vect{w}) =
	\begin{bmatrix}
		h_o(x_1,y_1,x_{c,i},y_{c,i},x_{a,i},y_{a,i},\alpha_i) \\
		h_o(x_2,y_2,x_{c,i},y_{c,i},x_{a,i},y_{a,i},\alpha_i) \\
		\vdots \\
		h_o(x_{N_p},y_{N_p},x_{c,i},y_{c,i},x_{a,i},y_{a,i},\alpha_i)
	\end{bmatrix}
	\leq \vect{0}.
\end{equation}

Moving obstacles are handled in a similar fashion, but letting the ellipsis center position and angle be time varying.
Obstacles in stand-on situations should, however, not be included in the constraints, since the mid-level algorithm is supposed to stand on in such situations.
Moreover, if an obstacle has entered an emergency state, the obstacle is so close and behaving unpredictably that the mid-level algorithm should disregard it and leave it for the short-term layer.
Hence, for the $i$-th moving obstacle not in a stand-on or an emergency situation, we define the constraint
\begin{equation}\label{eq:mid-level-dynamic-obs}
	\vect{h}_{m_i}(\vect{w}) = \begin{bmatrix}
		h_o(x_1,y_1,x_{c,i,1},y_{c,i,1},x_{a,i},y_{a,i},\alpha_{i,1}) \\		
		\vdots \\
 		h_o(x_{N_p},y_{N_p},x_{c,i,N_p},y_{c,i,N_p},x_{a,i},y_{a,i},\alpha_{i,N_p})
    \end{bmatrix}
    \leq \vect{0},
\end{equation}
where $x_{c,i,k}, y_{c,i,k}$ and $\alpha_{i,k}$ denote the position and course of the $i$-th moving obstacle at time step $k$.

Given $D$ static obstacles and $M$ obstacles not in stand-on or emergency situations, we define the constraint
\begin{equation}
	\bs h_o(\bs w, \bs \xi) = \begin{bmatrix}
		\vect{h}_{s_1}(\vect{w}) \\
		\vdots \\
		\vect{h}_{s_D}(\vect{w}) \\
		\vect{h}_{m_1}(\vect{w}) \\
		\vdots \\
		\vect{h}_{m_M}(\vect{w})
	\end{bmatrix} + \begin{bmatrix} \bs 0 \\ \bs \xi \end{bmatrix},
\end{equation}
where we include slack variables $\bs \xi \geq \bs 0$ on the moving obstacle constraints as part of a homotopy scheme.
The reason for using homotopy is that \gls{nlp} solvers in general only finds local minima, and can have issues with moving an initial guess ``through'' obstacles.
Normally, this is not an issue, but for the mid-level algorithm the optimal solution can change drastically from one iteration to another.
This can for instance happen if an obstacle enters a head-on or give-way state, where the solution can be trapped on the wrong side of an obstacle.
In general, homotopy describes introducing an extra parameter which is iteratively adjusted in order to iteratively move a local solution towards a global solution \citep{Deuflhard2011}.
In our homotopy scheme, we introduce slack variables on the moving obstacle constraints, which will allow solutions to travel through obstacles at the cost of a homotopy cost \eqref{eq:homotopy_cost} scaled by the homotopy parameter $K_\xi$.
Initially, this is selected as a low value to have a high amount of slack on the moving obstacles, while it is iteratively increased towards $K_\xi \to \infty$, which results in $\bs \xi = \bs 0$ and hence no slack on moving obstacles.
Currently, we only introduce slack on moving obstacles, but slack should also be introduced to static obstacles if they are small enough for the algorithm to be able to pass on both sides, like e.g. rocks, navigational marks, etc.

Similarly as in \citep{Eriksen2017b,Bitar2019ecc}, we ensure steady-state feasible trajectories at each time step through a constraint $\bs h_{\bs x_{r,k}}(\bs x_{r,k}) \leq \bs 0 \in \real^4$, which captures the state constraint $\bs x_r \in \mathcal{X}_r$ at time step $k$.
To ensure stead-state feasibility for the entire prediction horizon, we define the constraint
\begin{equation}
	\bs h_{\bs x_r}(\bs w) = \begin{bmatrix} \bs h_{\bs x_{r,k}}(\bs x_{r,0}) \\ \bs h_{\bs x_{r,k}}(\bs x_{r,1}) \\ \vdots \\ \bs h_{\bs x_{r,k}}(\bs x_{r,N_p-1}) \end{bmatrix} \leq \bs 0.
\end{equation}
Finally, the inequality constraints are combined as
\begin{equation}
	\bs h(\bs w, \bs \xi) = \begin{bmatrix} \bs h_o(\bs w, \bs \xi) \\ \bs h_{\bs x_r}(\bs w) \end{bmatrix} \in \real^{(M+D+4)N_p}.
\end{equation}

\section{Short-term COLAV}
	\label{sec:short-term}
For the short-term layer, the \acrfull{bc-mpc} algorithm is used, which is a sample-based \gls{mpc} algorithm intended for short-term \gls{asv} \gls{colav}.
The \gls{bc-mpc} algorithm was initially developed in \citep{Eriksen2019}, extended to also consider static obstacles in \citep{Eriksen2019b} and is experimentally validated in several full-scale experiments using a radar-based system for detecting and tracking obstacles.
The algorithm complies with \gls{colregs} rules 8, 13 and the second part of Rule 17, while favoring maneuvers complying with the maneuvering aspects of rules 14 and 15.
Notice that Rule 17 allows a ship to ignore the maneuvering aspects of rules 14 and 15 in situations where the give-way vessel does not maneuver.
The obstacle clearance will be larger if the algorithm ignores the maneuvering aspects of rules 14 and 15, like e.g. passing in front of an obstacle in a crossing situation where the ownship is the give-way vessel.
Moving obstacles are in general handled by the mid-level algorithm, making this applicable only in emergency situations and for obstacles detected so late that the mid-level algorithm is unable to avoid them.

The algorithm constructs a search space consisting of a finite number of trajectories, which each contain a sequence of maneuvers.
The maneuvers are constructed using a dynamic model of the ownship and a set of acceleration motion primitives, resulting in feasible trajectories being specified to the vessel controller.
For each maneuver, a discrete set of \gls{sog} and course accelerations are created as
\begin{equation}\label{eq:bc-mpc-acceleration-samples}
	\begin{aligned}
		\dot{\bs U}_{\text{samples}} &= \left\{ \dot U_1, \dot U_2, \ldots, \dot U_{N_U} \right\} \\
		\ddot{\bs \chi}_{\text{samples}} &= \left\{ \ddot \chi_1, \ddot \chi_2, \ldots, \ddot \chi_{N_\chi} \right\},
	\end{aligned}
\end{equation}
where $\dot U_i, i\in[1,N_U]$ and $\ddot \chi_i, i\in [1,N_\chi]$ denote $N_U\in\nat$ and $N_\chi\in\nat$ vessel-feasible speed and course accelerations.
Given the acceleration samples \eqref{eq:bc-mpc-acceleration-samples} and motion primitives for each maneuver in a trajectory, we create a set of desired \gls{sog} and course trajectories $\mathcal{U}_d$.
These trajectories have continuous acceleration, and is designed in an open-loop fashion by using the current reference tracked by the vessel controller for initialization, rather than the current vessel \gls{sog} and course.
The reason for this is that the reference to the vessel controller should be continuous in order to avoid jumps in the actuator commands.
To include feedback in the trajectory prediction, a set of feedback-corrected \gls{sog} and course trajectories $\bar{\mathcal{U}}_d$ is predicted using a simplified error model of the vessel and vessel controller.
Finally, the feedback-corrected \gls{sog} and course trajectories are used to compute a set of feedback-corrected pose trajectories:
\begin{equation}\label{eq:predictedPosition}
	\bar{\mathcal{H}} = \left\{\bar {\bs\eta}(\cdot) \big | (\bar U(\cdot), \bar \chi(\cdot)) \in \bar{\mathcal{U}} \right\},
\end{equation}
where $\bar {\bs\eta}(\cdot)$ denotes a kinematic simulation procedure that given \gls{sog} and course trajectories, $\bar U(\cdot)$ and $\bar \chi(\cdot)$, in $\bar{\mathcal{U}}_d$ computes the vessel pose.
See \citep{Eriksen2019,Eriksen2019b} for more details on the trajectory generation procedure.

In order to converge towards the trajectory specified by the mid-level algorithm, a desired acceleration is computed based on a \acrlong{los} guidance scheme.
In \citep{Eriksen2019} and \citep{Eriksen2019b}, the samples closest to the desired acceleration in \eqref{eq:bc-mpc-acceleration-samples} are replaced with the desired acceleration, given that this is vessel-feasible.
A problem with this, is that when operating at high speeds, the possible acceleration may not be symmetric, resulting in that zero acceleration (hence keeping a constant speed and course), may not be part of the search space.
This can cause undesirable behavior, since the \gls{bc-mpc} algorithm will be unable to keep the speed and course constant, which can cause oscillatory behavior.
In this paper, we therefore propose to move the acceleration samples closest to zero, and adding the desired acceleration as a separate sample, given that it is vessel feasible.
This will make sure that keeping a constant speed and course, as well as a trajectory converging towards the desired trajectory is included in the search space.

Given the predicted trajectories, the algorithm finds the optimal desired \gls{sog} and course trajectory for the vessel controller $\bs u_d^*(\cdot) = [U_d(\cdot)^*,\chi_d(\cdot)^*]$ as
\begin{equation}\label{eq:bc-mpc-ocp}
	\bs u_d^*(\cdot) = \underset{(\bar{\bs \eta}_k(\cdot), \bs u_{d,k}(\cdot)) \in (\bar{\mathcal{H}}, \mathcal{U}_d)}{\text{argmin}} G(\bar{\bs \eta}_k(\cdot), \bs u_{d,k}(\cdot); \bs p_d^{\text{mid}}(\cdot)),
\end{equation}
where the objective function is given as
\begin{multline}\label{eq:bc-mpc-objFunc}
 	G(\bar{\bs \eta}(\cdot), \bs u_d(\cdot);\bs p_d^{\text{mid}}(\cdot)) = w_{\text{al}}\text{align}(\bar{\bs \eta}(\cdot); \bs p_d^{\text{mid}}(\cdot)) + w_{\text{av,m}}\text{avoid}_\text{m}(\bar{\bs \eta}(\cdot)) + w_{\text{av,s}}\text{avoid}_\text{s}(\bar{\bs \eta}(\cdot)) \\ + w_{\text{t},U}\text{tran}_U(\bs u_d(\cdot)) + w_{\text{t},\chi}\text{tran}_\chi(\bs u_d(\cdot)).
\end{multline}
The variables $w_{\text{al}},w_{\text{av,m}},w_{\text{av,s}},w_{\text{t},U},w_{\text{t},\chi}>0$ are tuning parameters, while $\text{align}(\bar{\bs \eta}(\cdot); \bs p_d^{\text{mid}}(\cdot))$ measures the alignment between a candidate trajectory $\bar{\bs \eta}(\cdot)$ and the desired trajectory from the mid-level algorithm $\bs p_d^{\text{mid}}(\cdot)$.
The function $\text{avoid}_\text{m}(\bar{\bs \eta}(\cdot))$ ensures \gls{colav} of moving obstacles by penalizing trajectories close to obstacles, using a non-symmetric obstacle ship domain designed with the \gls{colregs} in mind.
The function $\text{avoid}_\text{s}(\bar{\bs \eta}(\cdot))$ ensures \gls{colav} of static obstacles by introducing an occupancy grid, while $\text{tran}_U(\bs u_d(\cdot))$ and $\text{tran}_\chi(\bs u_d(\cdot))$ introduces transitional costs to avoid shattering.
The transitional terms penalize deviations from the planned trajectory of the previous iteration, unless changing to the trajectory corresponding by the desired acceleration.
See \citep{Eriksen2019} and \citep{Eriksen2019b} for more details and descriptions of the terms.

\section{Simulation results}
	\label{sec:results}
The hybrid \gls{colav} system is verified through simulations, which are present in this section.
The simulations include ocean current and both static and moving obstacles.
We include moving obstacles both acting in compliance with the \gls{colregs}, and violating the \gls{colregs}.

The simulations are performed on a computer with an $2.8$ GHz Intel Core i7 processor, running macOS Mojave.

\subsection{Simulation setup}
The simulations are performed in MATLAB using CasADi \citep{Andersson2019} and IPOPT \citep{Wachter2005} for implementing the high-level and mid-level algorithms.
The simulator is built upon the mathematical model of the Telemetron \gls{asv} described in \autoref{sec:modeling}, and the model-based speed and course controller in \citep{Eriksen2018b} is used as the vessel controller.

The parameters of the high-level algorithm are listed in \autoref{tab:highlevelParam}.
\begin{table}
\caption{\label{tab:highlevelParam}Tuning parameters for the high-level algorithm.}
\begin{tabulary}{\linewidth}{llL}
\toprule
Param. & Value              & Comment \\
\midrule
$t_{\text{max}}$          &                                   & Maximum trajectory time \\
\hspace{.5cm}Scenario 1   & \SI{1420}{\second}                &  \\
\hspace{.5cm}Scenario 2   & \SI{1420}{\second}                &  \\
\hspace{.5cm}Scenario 3   & \SI{725}{\second}                 &  \\
$N_{\text{hi}}$           & \SI{1000}{}                       & Number of prediction steps \\
$K_e$                     & \SI{1.0}{\second\cubed\per\meter} & Energy penalty gain \\
$K_\delta$                & \SI{1.0}{}                        & Quadratic yaw control penalty gain \\
$L_m$                     & \SI{4.0}{\meter}                  & Length between control origin and outboard motor \\
\bottomrule
\end{tabulary}
\end{table}
The mid-level algorithm is implemented using the parameters in \autoref{tab:midlevelParam}.
\newcommand{\extraTabSpacing}{.2ex}
\begin{table}
\caption{\label{tab:midlevelParam}Tuning parameters for the mid-level algorithm.}
\begin{tabulary}{\linewidth}{llL}
\toprule
Param. & Value              & Comment \\
\midrule
$\overline{d}_{\text{CPA}}^{\text{enter}}$ & $900~\si{\meter}$  & State machine $d_{\text{CPA}}$ entry criteria \\\addlinespace[\extraTabSpacing]
$\underline{d}_{\text{CPA}}^{\text{exit}}$ & $2000~\si{\meter}$ & State machine $d_{\text{CPA}}$ exit criteria\\\addlinespace[\extraTabSpacing]
$[\underline{t}_{\text{CPA}}^{\text{enter}}, \overline{t}_{\text{CPA}}^{\text{enter}}]$ & $[0, 270]~\si{\second}$ & State machine $t_{\text{CPA}}$ entry criteria\\\addlinespace[\extraTabSpacing]
$[\underline{t}_{\text{CPA}}^{\text{exit}}, \overline{t}_{\text{CPA}}^{\text{exit}}]$ & $[-20, 290]~\si{\second}$ & State machine $t_{\text{CPA}}$ exit criteria\\\addlinespace[\extraTabSpacing]
$\overline{t}_{\text{crit}}^{\text{EM,enter}}$ & $20~\si{\second}$ & Emergency state $t_{\text{crit}}$ entry criteria\\\addlinespace[\extraTabSpacing]
$\underline{t}_{\text{crit}}^{\text{EM,exit}}$ & $25~\si{\second}$ & Emergency state $t_{\text{crit}}$ exit criteria\\\addlinespace[\extraTabSpacing]
$h$                 & \SI{10}{\second}      & Step size \\\addlinespace[\extraTabSpacing]
${N_p}$               & $36$            & Number of prediction steps \\\addlinespace[\extraTabSpacing]
$K_p$               & $0.02$     & Position error scaling \\\addlinespace[\extraTabSpacing]
$\sigma$               & $1$     & Huber loss function threshold \\\addlinespace[\extraTabSpacing]
$K_{\dot U}$            & $0.3$           & SOG-derivative penalty term scaling \\\addlinespace[\extraTabSpacing]
$K_{\dot \chi}$           & $2.5$           & Course-derivative penalty term scaling \\\addlinespace[\extraTabSpacing]
$K_{\text{HO}}$           & $40$           & Head-on potential function scaling \\\addlinespace[\extraTabSpacing]
$[\alpha_{x,\text{HO}}, \alpha_{y,\text{HO}}]$           & $[\sfrac{1}{500}, \sfrac{1}{400}]$           & Head-on potential function steepness parameters \\\addlinespace[\extraTabSpacing]
$x_{0,\text{HO}}$ & $1000~\si{\meter}$ & Head-on potential function attenuation parameter \\\addlinespace[\extraTabSpacing]
$K_{\text{GW}}$           & $40$           & Give-way potential function scaling \\\addlinespace[\extraTabSpacing]
$[\alpha_{x,\text{GW}}, \alpha_{y,\text{GW}}]$           & $[\sfrac{1}{400}, \sfrac{1}{500}]$           & Give-way potential function steepness parameters \\\addlinespace[\extraTabSpacing]
$y_{0,\text{GW}}$ & $-500~\si{\meter}$ & Give-way potential function attenuation parameter \\\addlinespace[\extraTabSpacing]
$K_{\text{SO}}$           & $3$           & Stand-on function scaling \\\addlinespace[\extraTabSpacing]
$K_{\text{EM}}$           & $3$           & Emergency function scaling \\\addlinespace[\extraTabSpacing]
$K_{\xi}$           & $[0.1, 1, 10, 100, \infty]$           & Iterative slack variable cost \\\addlinespace[\extraTabSpacing]
$x_a$           & $600~\si{\meter}$           & Moving obstacle ellipsis major axis size \\\addlinespace[\extraTabSpacing]
$y_a$           & $225~\si{\meter}$           & Moving obstacle ellipsis minor axis size \\
\bottomrule
\end{tabulary}  
\end{table}
The slack variable cost $K_\xi$ has five elements, implying that we use five steps in our homotopy scheme.
The mid-level \gls{nlp} is initially warm started with the solution from the previous iteration, while each step in the homotopy scheme is warm started with the solution from the previous step of the homotopy scheme, converging towards the solution without slack on the constraints.
To reduce the computational load and increase the predictability of the mid-level algorithm, we utilize six steps of each planned mid-level trajectory, only running the mid-level algorithm every $60~\si{\second}$.
This implies that six steps of the predicted solution will be implemented before computing a new solution, which further implies that the state machine is also only run every $60~\si{\second}$.
If the mid-level algorithm fails in finding a feasible solution, the algorithm will re-use the solution from the last iteration.
This may for instance happen if the algorithm tries to compute a solution while being inside a moving obstacle ellipse, which sometimes can be the case when an obstacle is exiting an emergency or stand-on state.
The \gls{bc-mpc} algorithm is run every $5~\si{\second}$, with parameters as described in \citep{Eriksen2019b}.
Static obstacles are padded with a safety margin of $150~\si{\meter}$ for the high-level and mid-level algorithms, while the \gls{bc-mpc} algorithm uses a safety margin of $100~\si{\meter}$ for static obstacles.
The reason for having a smaller static obstacle safety margin for the \gls{bc-mpc} algorithm is that it tends to struggle with following trajectories on the static obstacle boundaries.
The \gls{bc-mpc} algorithm would hence not be able to follow the nominal trajectory if the static obstacle safety margin was the same as for the mid-level and high-level algorithms.

The simulations are performed without any noise on the obstacle estimates, providing the algorithms with exact information about the obstacles position, course and speed.
The \gls{bc-mpc} algorithm has previously been shown to perform well with noisy and uncertain obstacle estimates in full-scale experiments using radar-based detection and tracking of obstacles \citep{Eriksen2019,Eriksen2019b}.
The mid-level algorithm is likely to have a larger requirement to low noise levels on the obstacle estimates, since the state machine in the mid-level algorithm depends on logic and discrete switching.
However, the algorithm is also run less frequently, reducing the required bandwidth of the obstacle estimates, possibly allowing using smoothing or tracking filters with a lower process noise if necessary.
It may also be feasible to make the mid-level algorithm depend on data from the \acrlong{ais}, which typically have much lower noise levels than radar-based tracking systems, while being subject to robustness issues \citep{Harati-Mokhtari2007}.

We present three scenarios, which demonstrate different important properties of the hybrid \gls{colav} system:
\begin{desclist}{\hspace{.3cm}\bf}{ }[Scenario 2]
  \item[Scenario 1] This scenario contains two static obstacles, and four moving obstacles of which all comply with the \gls{colregs}.
  The moving obstacles demonstrate stand-on, give-way and head-on situations.
  \item[Scenario 2] This scenario contains one static and five moving obstacles.
  The moving obstacles demonstrate stand-on with an obstacle ignoring the \gls{colregs}, an overtaking and a simultaneous head-on, give-way and stand-on situation with obstacles complying with the \gls{colregs}.
  \item[Scenario 3] This scenario contains two moving obstacles, which suddenly perform dangerous maneuvers close to the ownship, displaying the use of the emergency state.
\end{desclist}

\subsection{Scenario 1}
\begin{figure}
	\centering
	\begin{tabular}[b]{@{}c@{}}
		\includegraphics[width=.7\linewidth]{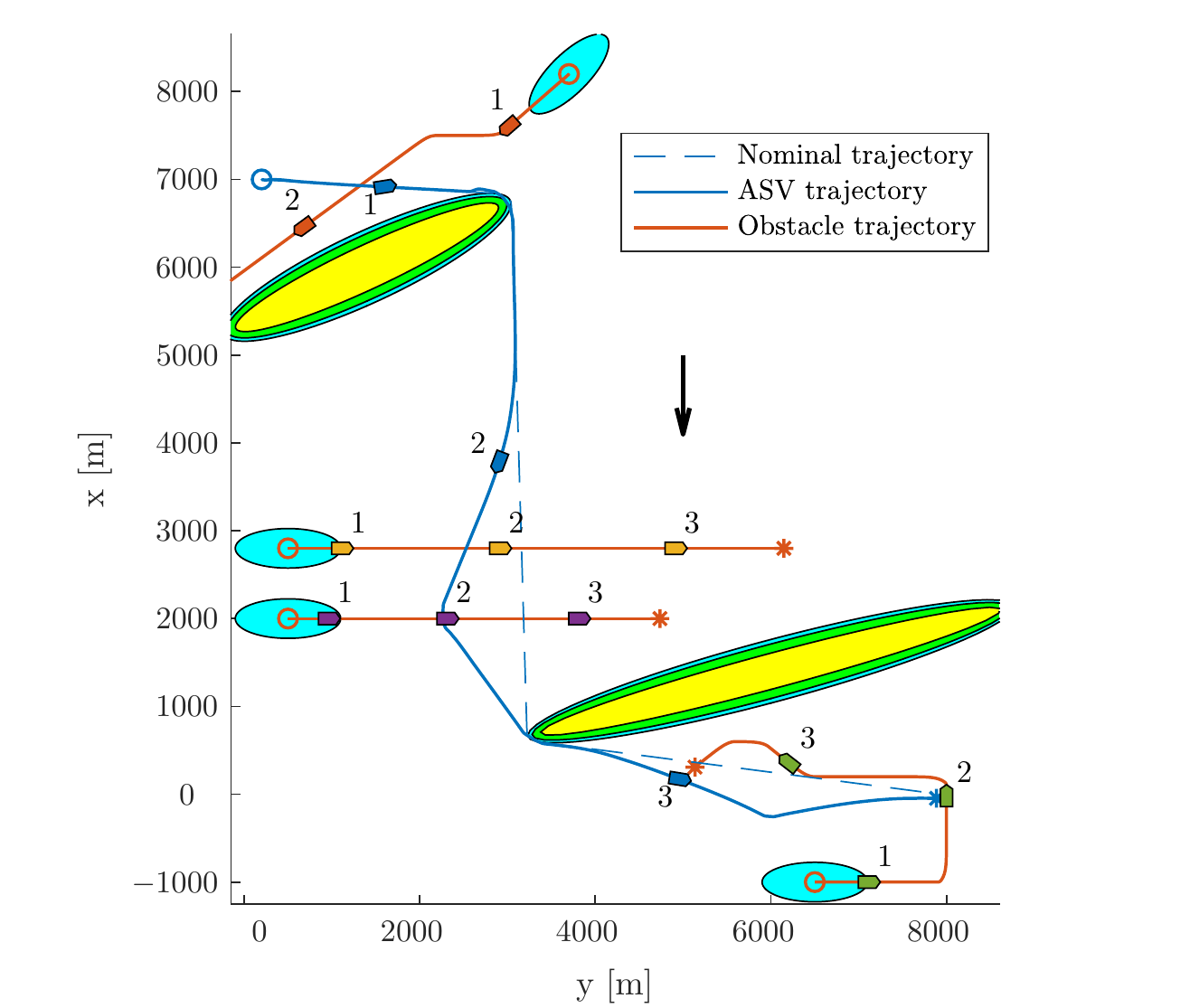}\\
		  \subfigCaption{A}{subfig:scenario_1_trajectory}{Trajectory plot. 
		  The initial position of the ownship and obstacles are shown with circles, with the blue ellipses illustrating the moving obstacle ellipse size.
		  The vessel patches, which are overexaggerated for visualization, mark the ownship and obstacle poses at given time stamps.
		  The static obstacles are shown in yellow, with the \acrshort{bc-mpc} and mid-level safety margins enclosed around.
      The black arrow indicates the ocean current direction.} \\
  	\includegraphics[width=.7\linewidth]{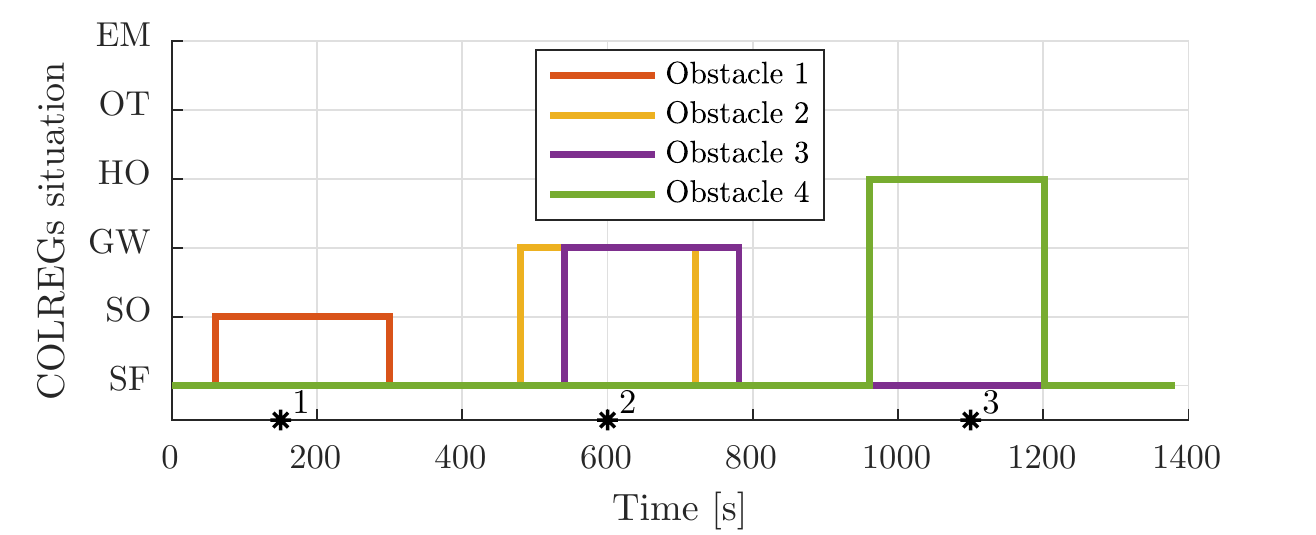} \\
      \subfigCaption{B}{subfig:scenario_1_COLREGs}{Output from the state machine for each obstacle.
      The asterisks mark time stamps, and the colors correspond to the obstacle patch colors in the trajectory plot.}
	\end{tabular}
    \caption{\label{fig:scenario_1_trajectory}Scenario 1: Trajectory and COLREGs interpretation.
    The text marks denote the time steps $[150, 600, 1100]~\si{\second}$.}
\end{figure}
Scenario 1 contains two static obstacles, four moving obstacles, an ocean current of $[-2,0]\tr~\si{\meter\per\second}$ and is shown in \autoref{fig:scenario_1_trajectory}.
The high-level planner plans a nominal trajectory between the initial and goal positions at $[7000,200]\tr~\si{\meter}$ and $[0,7900]\tr~\si{\meter}$, respectively.
The first obstacle is in a stand-on situation, where it is required to maneuver in order to avoid collision with the ownship, which is required to stand on.
As shown in \ref{subfig:scenario_1_COLREGs}, the first obstacle is quickly considered as a stand-on situation, at which the mid-level algorithm disregards the obstacle and continues with the current speed and course.  
Following this, the obstacle maneuvers in accordance to the \gls{colregs}, and we avoid collision.
After the first static obstacle, we encounter two crossing vessels where the ownship is deemed the give-way vessel.
In accordance with the \gls{colregs}, we maneuver to starboard in order to pass behind both obstacles.
Notice that the second give-way obstacle is detected as a give-way situation later than the first, since the entry criteria in the state machine includes the time to \gls{cpa}, which is higher for the second give-way obstacle.
After avoiding the two give-way obstacles, we converge towards the nominal trajectory and encounter a head-on situation.
This is correctly identified by the state machine as head on, and we maneuver to starboard in order to avoid collision.
Notice that even though the obstacle maneuvers, we keep the obstacle in the head-on state until we have passed it.
\begin{figure}
  	\centering
  	\begin{tabular}[b]{@{}c@{}}
      	\includegraphics[width=.7\linewidth]{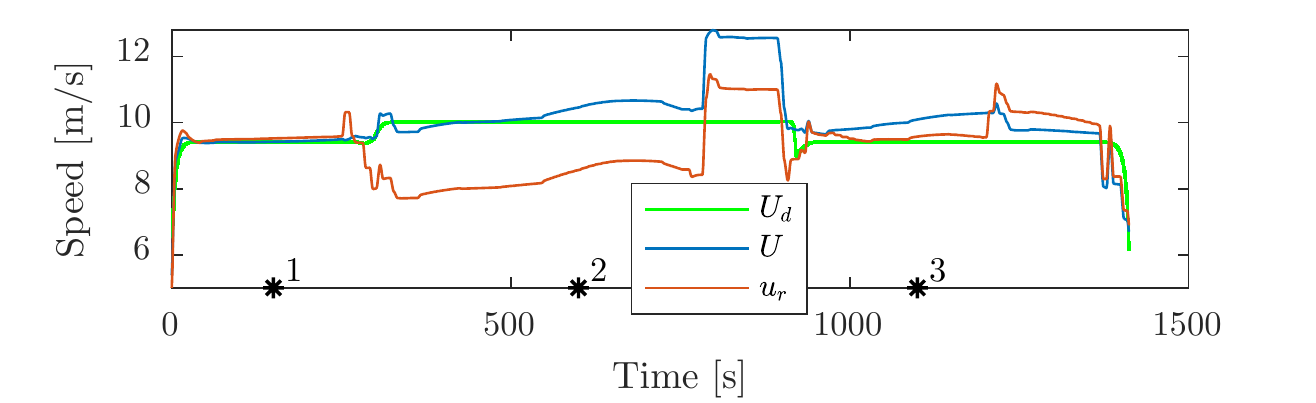} \\
      	\subfigCaption{A}{subfig:scenario_1_speed}{Speed trajectories}\\
      	\includegraphics[width=.7\linewidth]{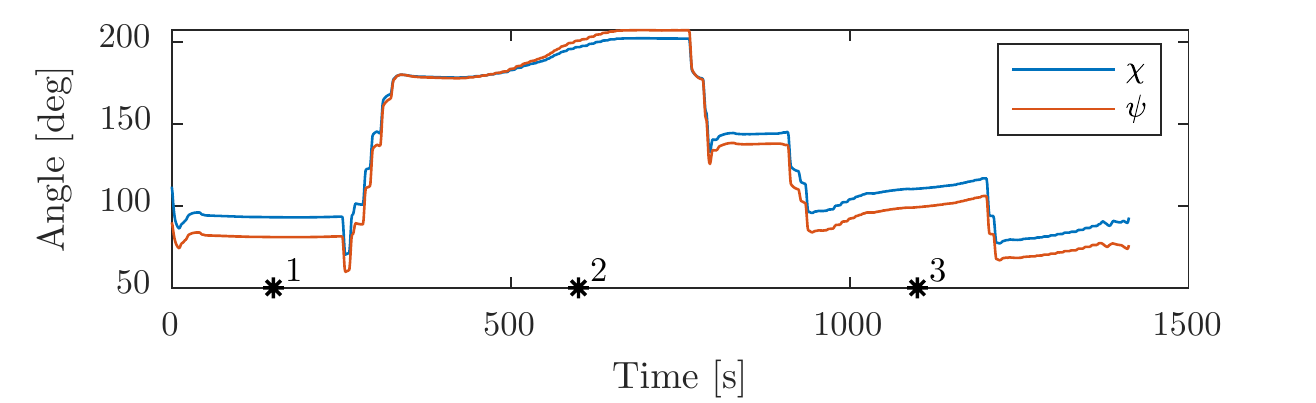} \\
      	\subfigCaption{B}{subfig:scenario_1_angles}{Angular trajectories}
  	\end{tabular}
    \caption{\label{fig:scenario_1_speed}Scenario 1: Speed and angular trajectories. 
    The asterisks mark the same time samples as in \autoref{fig:scenario_1_trajectory}.}
\end{figure}
\autoref{fig:scenario_1_speed} shows the speed and angular trajectories during Scenario 1, where the desired speed is calculated as the nominal speed at the closest point on the nominal trajectory given the ownship position.
From this, we see that the mid-level and \gls{bc-mpc} algorithms manage to track the desired nominal speed before and after the first static obstacle, where no obstacles require maneuvering away from the nominal trajectory.
Notice that when encountering the two crossing obstacles, the mid-level algorithm chooses to slowly change the course, which is due to the attenuation of the give-way potential function and the large distance between the vessels.
It would be better to make a clear course change, which is a subject of tuning.
After passing the two crossing obstacles, the mid-level algorithm increases the speed in order to get back to the nominal trajectory, which is due to the algorithm attempting to keep the speed projected on the nominal trajectory equal as the desired nominal speed.
Furthermore, notice that the mid-level algorithm actively controls the relative surge speed in order achieve the desired \gls{sog}, which is clearly seen when passing the first static obstacle.

\subsection{Scenario 2}
\begin{figure}
	\centering
	\begin{tabular}[b]{@{}c@{}}
		\includegraphics[width=.7\linewidth]{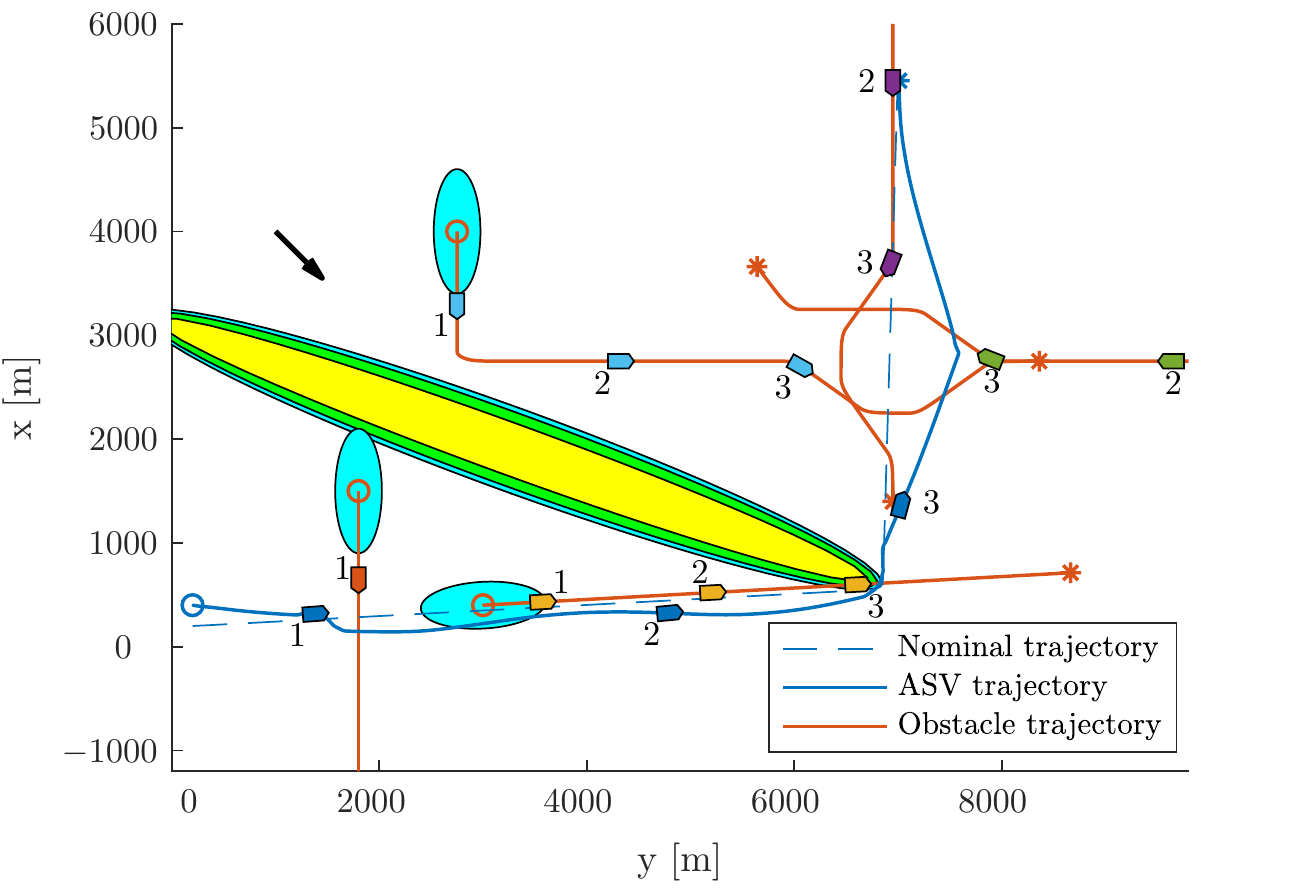}\\
		  \subfigCaption{A}{subfig:scenario_2_trajectory}{Trajectory plot. 
		  The initial position of the ownship and obstacles are shown with circles, with the blue ellipses illustrating the moving obstacle ellipse size.
		  The vessel patches, which are overexaggerated for visualization, mark the ownship and obstacle poses at given time stamps.
		  The static obstacles are shown in yellow, with the \acrshort{bc-mpc} and mid-level safety margins enclosed around.
      The black arrow indicates the ocean current direction.} \\
  	\includegraphics[width=.7\linewidth]{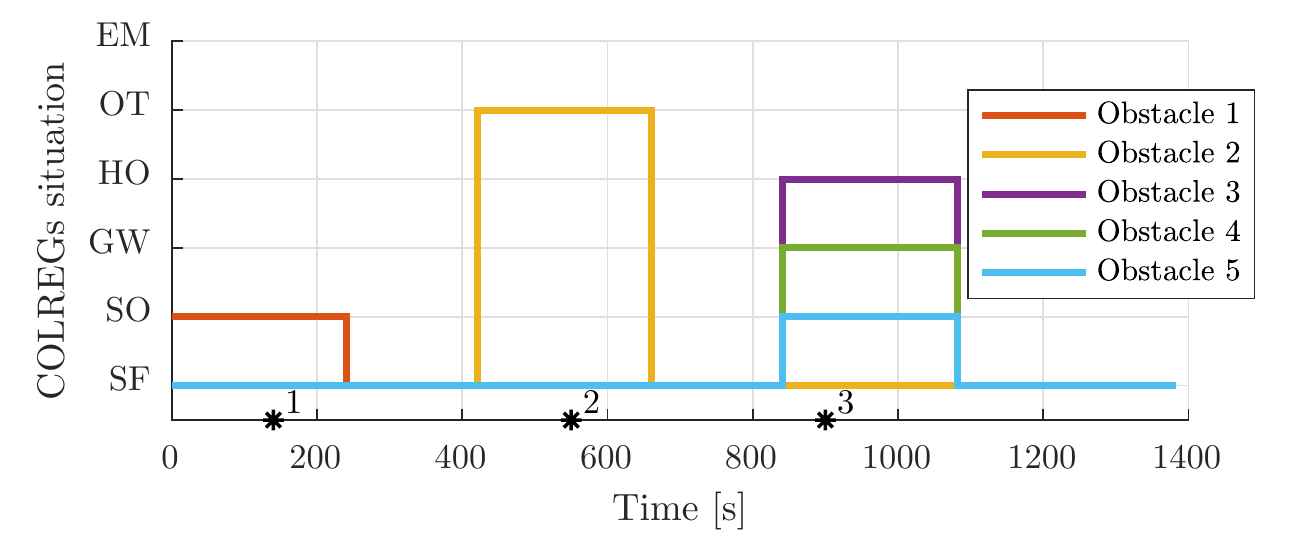} \\
      \subfigCaption{B}{subfig:scenario_2_COLREGs}{Output from the state machine for each obstacle.
      The asterisks mark time stamps, and the colors correspond to the obstacle patch colors in the trajectory plot.}
	\end{tabular}
    \caption{\label{fig:scenario_2_trajectory}Scenario 2: Trajectory and COLREGs interpretation.
    The text marks denote the time steps $[140, 550, 900]~\si{\second}$.}
\end{figure}
Scenario 2, shown in \autoref{fig:scenario_2_trajectory}, is more complex than Scenario 1, with a total of five moving obstacles, and has an ocean current of $[-1,1]\tr~\si{\meter\per\second}$.
The high-level planner plans a nominal trajectory between the initial and goal positions at $[200,200]\tr~\si{\meter}$ and $[5500,7000]\tr~\si{\meter}$, respectively.
The first obstacle is a crossing vessel, which similarly as in Scenario 1 is deemed to give way for the ownship, which should keep the current speed and course.
However, in this scenario, the obstacle violates the \gls{colregs} by not maneuvering in order to avoid collision.
Therefore, the \gls{bc-mpc} algorithm maneuvers to avoid collision when the obstacle gets so close that the safety margins of the \gls{bc-mpc} algorithms is violated.
The \gls{bc-mpc} algorithm maneuvers to port, as advised by \gls{colregs} Rule 17 for crossing situations where the stand-on vessel has to maneuver, and safely avoid the first obstacle.
The second obstacle is overtaken by the ownship, and correctly considered as an overtaking situation by the state machine.
For such an situation, there is no requirement on how the ownship should maneuver, except keeping clear from the overtaken vessel.
After passing the second obstacle, we encounter a complex situation with simultaneous head-on, give-way and stand-on obligations.
In this situation, each vessel, including the ownship, finds itself in a situation where a head-on and a give-way situation require starboard maneuvers, while a stand-on situation requires the vessel to keep the current speed and course.
However, head-on and give-way obligations should be prioritized higher than stand-on situations, and the situation is quite easily solved by each vessel maneuvering to starboard and passing behind the vessel crossing from starboard.
The mid-level algorithm solves this situation with the desirable behavior, and converges towards the nominal trajectory after the situation is resolved.
As shown in \ref{subfig:scenario_2_COLREGs}, the state machine interprets the situations correctly.
\begin{figure}
  	\centering
  	\begin{tabular}[b]{@{}c@{}}
      	\includegraphics[width=.7\linewidth]{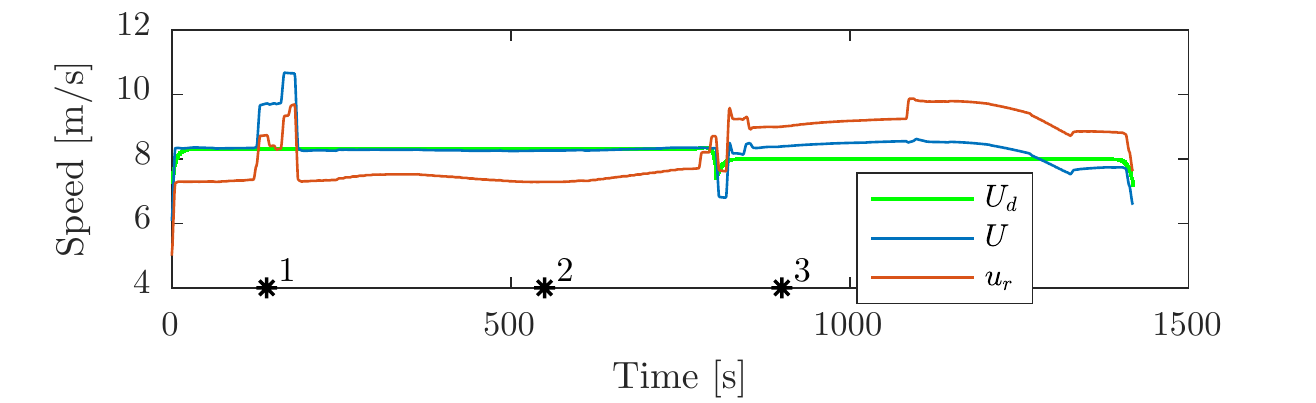} \\
      	\subfigCaption{A}{subfig:scenario_2_speed}{Speed trajectories}\\
      	\includegraphics[width=.7\linewidth]{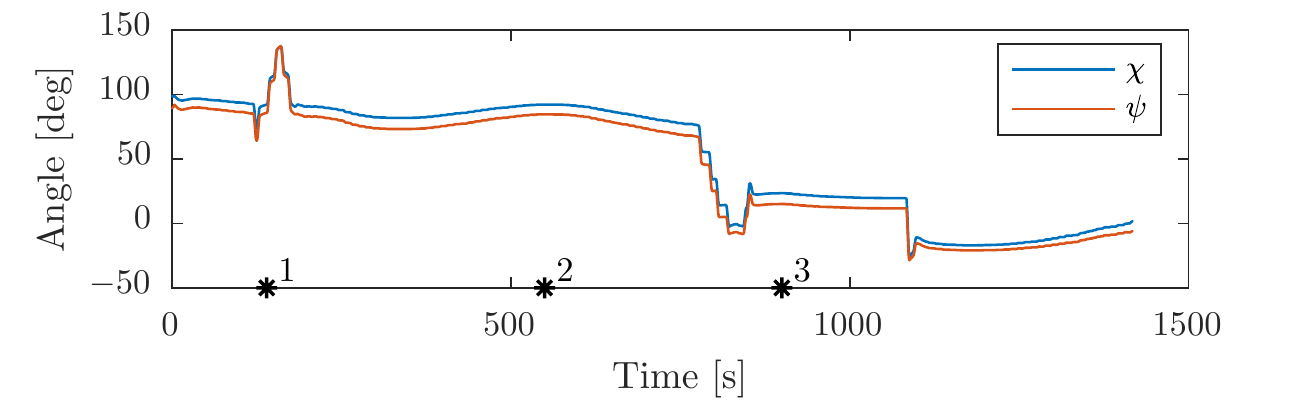} \\
      	\subfigCaption{B}{subfig:scenario_2_angles}{Angular trajectories}
  	\end{tabular}
    \caption{\label{fig:scenario_2_speed}Scenario 2: Speed and angular trajectories. 
    The asterisks mark the same time samples as in \autoref{fig:scenario_2_trajectory}.}
\end{figure}
From the speed trajectory in \autoref{fig:scenario_2_speed} it is clear that the mid-level algorithm follows the desired nominal speed also when overtaking the second obstacle.

\subsection{Scenario 3}
\begin{figure}
	\centering
	\begin{tabular}[b]{@{}c@{}}
		\includegraphics[width=.7\linewidth]{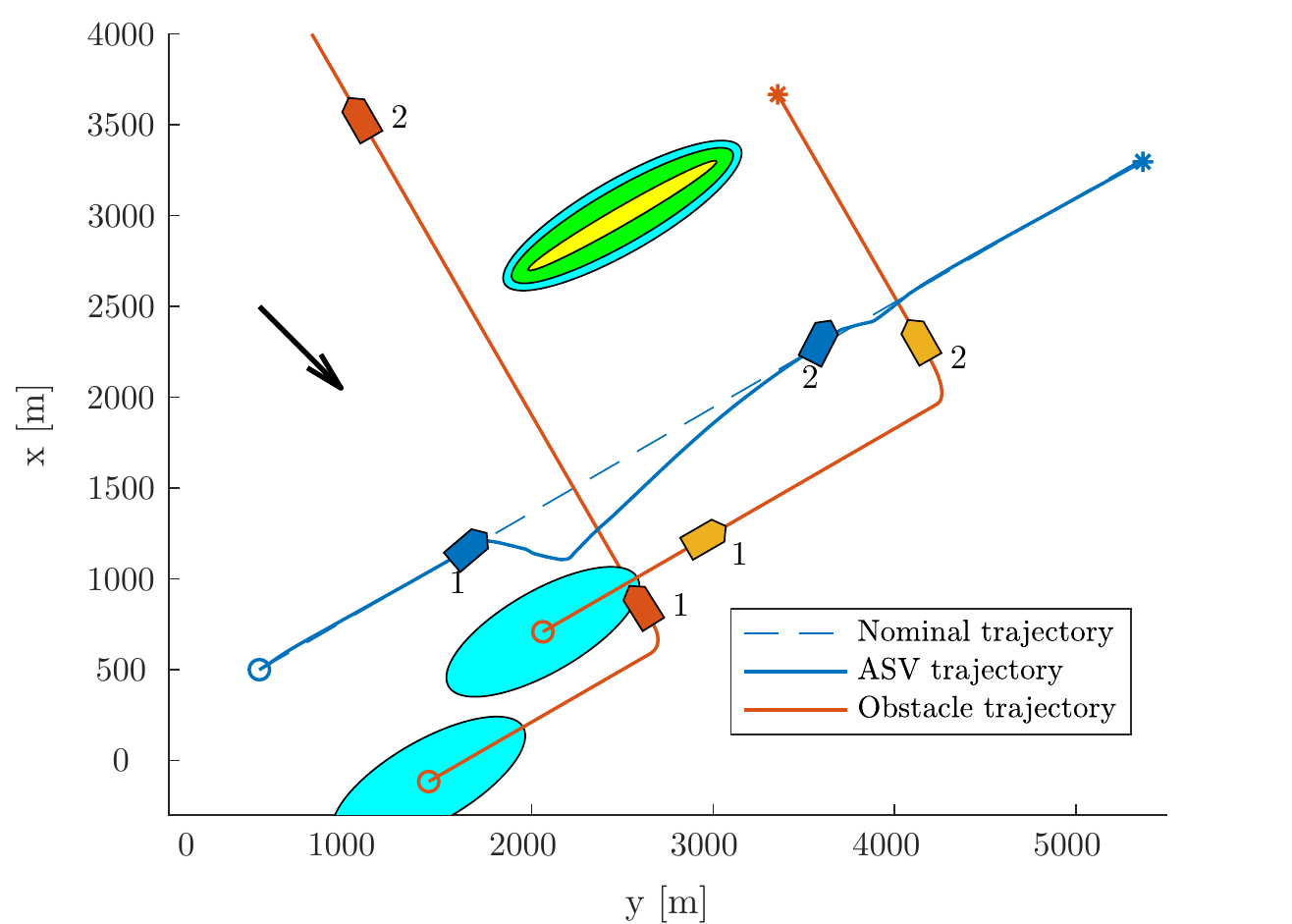}\\
		  \subfigCaption{A}{subfig:scenario_3_trajectory}{Trajectory plot. 
		  The initial position of the ownship and obstacles are shown with circles, with the blue ellipses illustrating the moving obstacle ellipse size.
		  The vessel patches, which are overexaggerated for visualization, mark the ownship and obstacle poses at given time stamps.
		  The static obstacles are shown in yellow, with the \acrshort{bc-mpc} and mid-level safety margins enclosed around. 
      The black arrow indicates the ocean current direction.} \\
  	\includegraphics[width=.7\linewidth]{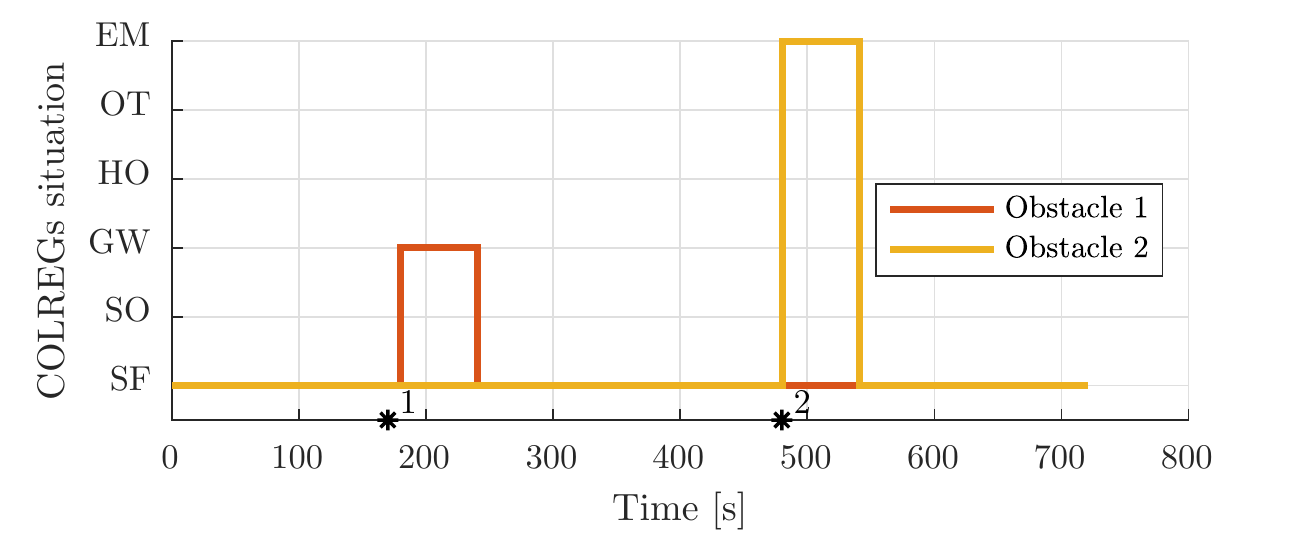} \\
    	\subfigCaption{B}{subfig:scenario_3_COLREGs}{Output from the state machine for each obstacle.
    	The asterisks mark time stamps, and the colors correspond to the obstacle patch colors in the trajectory plot.}
	\end{tabular}
    \caption{\label{fig:scenario_3_trajectory}Scenario 3: Trajectory and COLREGs interpretation.
    The text marks denote the time steps $[170, 480]~\si{\second}$.}
\end{figure}
Scenario 3, shown in \autoref{fig:scenario_3_trajectory}, contains two moving obstacles on parallel courses with the ownship, and has an ocean current of $[-1,1]\tr~\si{\meter\per\second}$.
The high-level planner plans a nominal trajectory between the initial and goal positions at $[500,500]\tr~\si{\meter}$ and $[3328,5399]\tr~\si{\meter}$, respectively, which results in a straight line trajectory with a course angle of $60\si{\degree}$.
The first obstacle travels at a higher speed than the ownship, while the second one travels at a lower speed and will be overtaken by the ownship.
Since the obstacles are on parallel paths with the obstacle, the time to \gls{cpa} is sufficiently high such that the obstacles are in the safe state, even though the the vessels are quite close.
However, both obstacles make sudden maneuvers to port dangerously close to the ownship and enters on a crossing course with the ownship.
With respect to the \gls{colregs}, the ownship is required to give way to both obstacles since they are crossing from the ownship's starboard side.
One can, however, argue that the maneuvers displayed by the obstacles are dangerous and displays poor seamanship, such that the ownship should not be held accountable if a collision occurred.
Nevertheless, the hybrid \gls{colav} system manages to avoid both obstacles.
As seen in \ref{subfig:scenario_3_COLREGs}, the first obstacle is sufficiently far away from the ownship to be considered as a give-way situation when the state machine interprets the situation, and the mid-level algorithm plans a trajectory passing behind the first obstacle.
The second obstacle maneuvers to port even closer to the ownship, resulting in the distance to the critical point being within the threshold for entering the emergency situation when the state machine interprets the situation.
In this situation, the mid-level algorithm disregards the obstacle and leaves it to the \gls{bc-mpc} algorithm to avoid collision.
\begin{figure}
  	\centering
  	\begin{tabular}[b]{@{}c@{}}
      	\includegraphics[width=.7\linewidth]{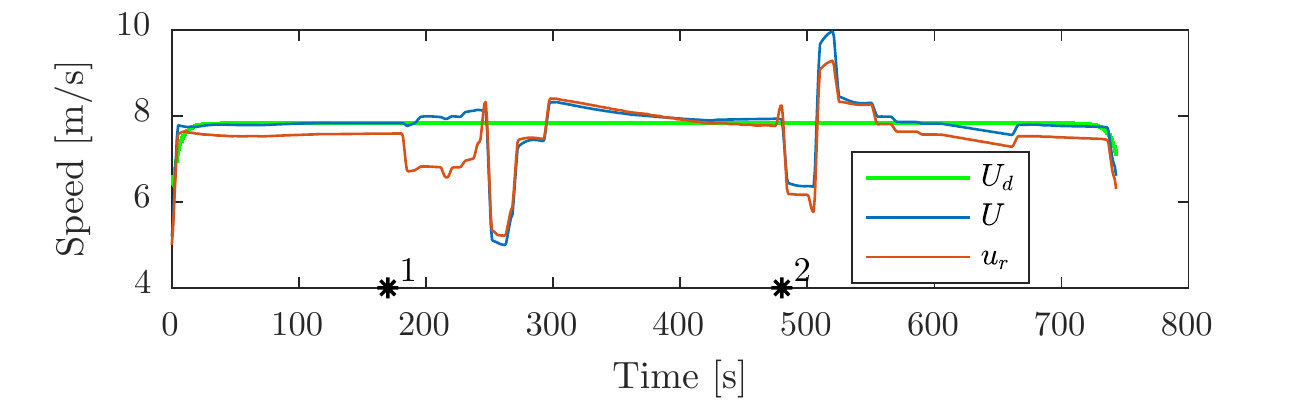} \\
      	\subfigCaption{A}{subfig:scenario_3_speed}{Speed trajectories}\\
      	\includegraphics[width=.7\linewidth]{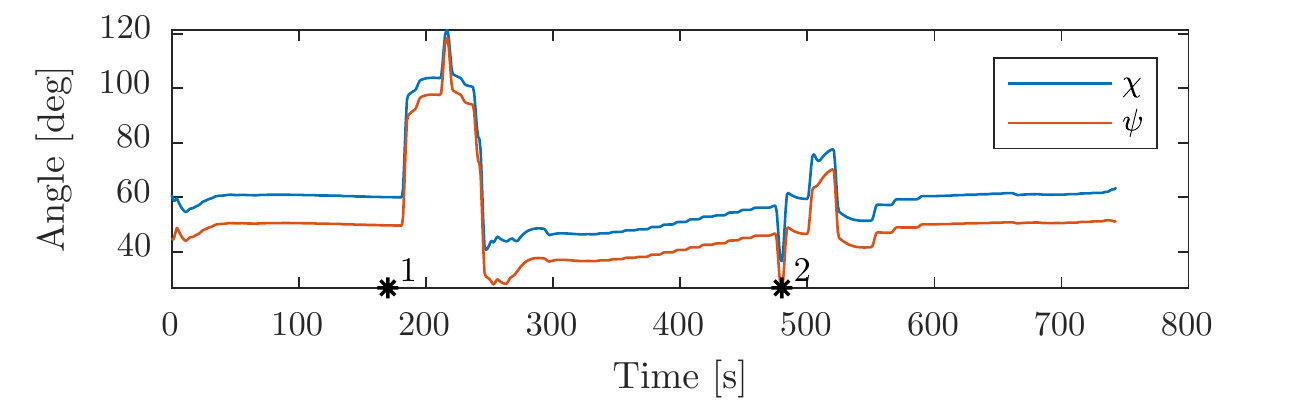} \\
      	\subfigCaption{B}{subfig:scenario_3_angles}{Angular trajectories}
  	\end{tabular}
    \caption{\label{fig:scenario_3_speed}Scenario 3: Speed and angular trajectories. 
    The asterisks mark the same time samples as in \autoref{fig:scenario_3_trajectory}.}
\end{figure}
As seen in \autoref{fig:scenario_3_speed}, the mid-level algorithm both reduces the speed and changes the course to avoid the first obstacle.
When approaching the second obstacle, the \gls{bc-mpc} algorithm initiates a speed reduction, and after some time also maneuver to starboard in order to pass behind the obstacle and resolve the situation.

\subsection{Simulation summary}
The simulation results show that the hybrid \gls{colav} system is able to handle a wide range of situations, while also behaving in an energy-optimal way when moving obstacles are not interfering with the ownship trajectory.
\autoref{tab:minimum_distances} shows the minimum distance to static and moving obstacles for the scenarios.
\begin{table}
  \caption{\label{tab:minimum_distances}Minimum distance to static and moving obstacles for the simulation scenarios.}
  \begin{tabular}{lllllll}
    \toprule
    \multirow{ 2}{*}{\textbf{Scenario}} & \multirow{ 2}{*}{\begin{tabular}{@{}c@{}} \textbf{Minimum distance to} \\ \textbf{static obstacles}\end{tabular}} & \multicolumn{5}{c}{\textbf{Minimum distance to moving obstacle number}} \\
    &  & \textbf{1} & \textbf{2} & \textbf{3} & \textbf{4} & \textbf{5} \\
    \midrule
    Scenario 1 & 93.7~\si{\meter} & $634.3~\si{\meter}$ & $596.3~\si{\meter}$ & $522.7~\si{\meter}$ & $726.8~\si{\meter}$ & -- \\
    Scenario 2 & 118.2~\si{\meter} & $185.5~\si{\meter}$ & $228.3~\si{\meter}$ & $1097.2~\si{\meter}$ & $575.6~\si{\meter}$ & $842.3~\si{\meter}$ \\
    Scenario 3 & 1123.8~\si{\meter} & $326.4~\si{\meter}$ & $106.6~\si{\meter}$ & -- & -- & -- \\
    \bottomrule
  \end{tabular}
\end{table}
The minimum distance to static obstacles is in Scenario 1 below the safety region size of the \gls{bc-mpc} algorithm, which is intentional and caused by the algorithm using a smooth penalty function for interpreting static obstacles.
The penalty function value increases linearly when moving further into the safety region, see \citep{Eriksen2019b} for more details.
The minimum distance to moving obstacles is a bit difficult to interpret, since the obstacle ship domains are non-circular, implying that the required clearance depends on relative position of the ownship with respect to the moving obstacles.
However, we see that we have a larger clearance in head-on, give-way and stand-on situations where the obstacles comply with the \gls{colregs}, and do not perform dangerous maneuvers (as in Scenario 3), compared to overtaking situations.
The reason for this is that when overtaking (obstacle 2 in Scenario 2), we pass the obstacle on a parallel course, resulting in the minor axis of the moving obstacle ellipsis indicating the required clearance.
Furthermore, we see that obstacle 1 in Scenario 2, which ignores its give-way obligation, comes significantly closer than other crossing obstacles except for those in Scenario 3.
The reason for this is that the \gls{bc-mpc} algorithm, which handles this situation, has a lower clearance requirement than the mid-level algorithm, which still should be considered as safe.
In Scenario 3, the two obstacles display poor seamanship, and behave dangerously.
Obstacle 1 is handled by the mid-level algorithm and passed with a clearance lower than the major axis of the mid-level algorithm, which is caused by the \gls{bc-mpc} algorithm ``cutting the corner''.
The clearance should still be considered safe since we are behind the obstacle, and the clearance requirements of the \gls{bc-mpc} algorithm is enforced.
Obstacle 2, which is placed in the emergency state and handled by the \gls{bc-mpc} algorithm, is passed with a clearance of only $106.6~\si{\meter}$. 
This is lower than the clearance to Obstacle 1 in Scenario 2 (which violated its stand-on requirement), and is due to the \gls{bc-mpc} algorithm having a non-symmetric obstacle ship domain function allowing for a smaller clearance when passing behind an obstacle than in front.

For the three scenarios, the high-level planner used an average of $67~\si{\second}$ with a maximum of $93~\si{\second}$ to compute the solution.
Since the high-level planner is intended to be run off-line, this is well within reasonable limits. 
The mid-level algorithm used $0.60~\si{\second}$ on average, and a maximum of $2.1~\si{\second}$, which we consider to be real-time feasible since the mid-level algorithm only is run every $60~\si{\second}$.
The \gls{bc-mpc} algorithm used $0.29~\si{\second}$ on average, and a maximum of $0.63~\si{\second}$, which we also consider to be real-time feasible when the \gls{bc-mpc} algorithm is run every $5~\si{\second}$.
The \gls{bc-mpc} algorithm is highly parallelizable, which could reduce the \gls{bc-mpc} runtime by a large magnitude if required.

\section{Conclusion}
	\label{sec:conclusion}
In this paper, we have presented a three-layered hybrid \gls{colav} system, compliant with \gls{colregs} rules 8 and 13--17.
As part of this, we have further developed the \gls{mpc}-based mid-level \gls{colav} algorithm in \citep{Eriksen2017b,Bitar2019ecc} to comply with \gls{colregs} rules 13--16 and parts of Rule 17, which includes developing a state machine for \gls{colregs} interpretation.
The hybrid \gls{colav} system has a  well-defined division of labor, including an inherent understanding of \gls{colregs} Rule 17, where the mid-level algorithm obeys stand-on situations, while the \gls{bc-mpc} algorithm handles situations where give-way vessels does not maneuver.

The hybrid \gls{colav} system is verified through simulations, where we in three scenarios challenge the system with a number of different situations.
The scenarios include multi-obstacle situations with multiple simultaneously active \gls{colregs} rules, and situations where obstacles violate the \gls{colregs}.
Collision is avoided in all the scenarios, and we show that the ownship follows an energy-optimized trajectory generated by the high-level planner when moving obstacles does not interfere with this trajectory.

For further work, we suggest to:
\begin{itemize}
	\item Investigate if using situation-dependent entry and exit criteria parameters in the state machine improves the performance.
	\item Expand the state machine with the possibility of transitioning from head-on, give-way and overtaking states to the emergency state for situations where obstacles behave dangerously or hostile.
	\item Develop a methodology for deciding tuning parameters.
	\item Perform simulations with noisy obstacle estimates to investigate how the state machine and mid-level algorithm respond to this.
	\item Explore the possibilities for integrating the \gls{colregs} interpretation in the mid-level \gls{nlp}, relaxing the assumption of the current \gls{colregs} situation being valid for the entire prediction horizon.
	\item Simulate scenarios where multiple vessels running the hybrid \gls{colav} system interact with each other.
	\item Validate the hybrid \gls{colav} system in full-scale experiments.
\end{itemize}

\section*{Conflict of Interest Statement}
The authors declare that the research was conducted in the absence of any commercial or financial relationships that could be construed as a potential conflict of interest.

\section*{Author Contributions}
The work in this article is the result of a collaboration between BE and GB, supervised by MB and AL.
The contributions to the mid-level and short-term algorithms are made by BE, while the \gls{colregs} interpreter is developed in collaboration between GB and BE.
GB has implemented the high-level planner, and prepared this for integration with the developed simulator.
BE has taken lead on writing the paper, in collaboration with GB.
MB and AL have provided valuable feedback in the writing process.

\section*{Funding}
This work was supported by the Research Council of Norway through project number 269116, project number 244116, as well as the Centers of Excellence funding scheme with project number 223254.

\bibliographystyle{frontiersinSCNS_ENG_HUMS} 
\bibliography{references}

\end{document}